\documentclass[twocolumn, nofootinbib, aps, prd, 10pt]{revtex4-1}
\usepackage{graphicx}
\usepackage{float}
\usepackage{url}
\usepackage{color}
\usepackage[dvipsnames]{xcolor}
\usepackage{rotating}
\usepackage{amssymb,amsmath}
\usepackage{mathtools} 
\usepackage{booktabs}
\usepackage{bbm}
\usepackage{bm}
\usepackage{enumerate}
\pdfoutput=1

\newcommand{\fsky}{f_{\rm sky}}

\newcommand{\aap}{{A\&A}}
\newcommand{\apjs}{{Astrophys.~J.~Supp.}}
\newcommand{\aj}{{Astro.~Journal}}

\newcommand{\mnras}{{Mon.~Not.~R.~Astron.~Soc.}}

\newcommand{\Ylm}{Y_{\ell m}}

\newcommand{\alm}{a_{\ell m}}

\newcommand{\slm}{s_{\ell m}}

\newcommand{\Clss}{\Cl^{ss}}
\newcommand{\Clsstilde}{\tilde{C}_\ell^{ss}}

\newcommand{\Cltt}{\Cl^{tt}}

\newcommand{\ClTT}{{\bf \Cl^{TT}}}

\newcommand{\bfTell}{{\bf T}_\ell}

\newcommand{\Cl}{C_\ell}

\newcommand{\Cltrue}{\Clsstilde}
\newcommand{\Clest}{\tilde{C}_{\ell}^{\rm est}}

\newcommand{\ellmax}  {\ell_{\mathrm{max}}}

\newcommand{\Npix}{N_{\mathrm{pix}}}
\newcommand{\Nobs}{N_{\mathrm{obs}}}
\newcommand{\Nsys}{N_{\mathrm{sys}}}
\newcommand{\fic}{s_0}
\newcommand{\deltaobs}{\delta_{\mathrm{obs}}}
\newcommand{\Ntrue}{N_{\mathrm{true}}}
\newcommand{\Nbar}{\bar{N}}
\newcommand{\pixavg}[1]{\langle#1\rangle_{\rm pix}}

\newcommand{\nhat}{{\bf \hat n}}
\newcommand{\shat}{\hat s}

\newcommand{\Ntpl}{N_{\rm tpl}}
\newcommand{\rhotpl}{\rho_{\rm tpl}}

\newcommand{\dchisqcl}{\Delta\chi^2_{\rm \Cl}}
\newcommand{\dchisqthresh}{\Delta\chi^2/\Delta\chi^2_0}

\newcommand{\hatalphamp}{\hat{\alpha}_{\rm mp}}

\newcommand{\dobs}{d_{\rm obs}}
\newcommand{\dadd}{d_{\rm add}}
\newcommand{\fsys}{f_{\rm sys}}

\newcommand{\hatfsys}{\hat{f}_{\rm sys}}

\newcommand{\balpha}{\pmb{\alpha}}

\newcommand{\varsys}{{\sigma^2_{\rm sys}}}
\newcommand{\varsig}{{\sigma^2_{\rm sig}}}
\newcommand{\vartt}{{\sigma^2_{tt}}}
\newcommand{\var}[1]{{\sigma^2_{#1}}}
\newcommand{\tilvar}[1]{{\tilde{\sigma}^2_{#1}}}

\newcommand{\sysvar}{{\varsys}}
\newcommand{\sigvar}{{\varsig}}

\everymath{\displaystyle}
  
\newcommand{\be}{\begin{equation}}
\newcommand{\ee}{\end{equation}}

\begin{document}

\title{Mitigating contamination in LSS surveys: a comparison of methods}

\author{Noah Weaverdyck$^*$ and Dragan Huterer}
\affiliation{Department of Physics, University of Michigan, 
450 Church St, Ann Arbor, MI 48109-1040}
\affiliation{Leinweber Center for Theoretical Physics, University of Michigan, 450 Church St, Ann Arbor, MI 48109-1040}
\email[Corresponding author: ]{nweaverd@umich.edu}
\date{\today}

\begin{abstract}
Future large scale structure surveys will measure the locations and shapes of billions of galaxies. The precision of such catalogs will require meticulous treatment of systematic contamination of the observed fields. We compare several existing methods for removing such systematics from galaxy clustering measurements. We show how all the methods, including the popular pseudo-$\Cl$ Mode Projection and Template Subtraction methods, can be interpreted under a common regression framework and use this to suggest improved estimators. We show how methods designed to mitigate systematics in the power spectrum can be used to produce clean maps, which are necessary for cosmological analyses beyond the power spectrum, and we extend current methods to treat the next-order multiplicative contamination in observed maps and power spectra. Two new mitigation methods are proposed, which incorporate desirable features of current state-of-the-art methods while being simpler to implement. Investigating the performance of all the methods on a common set of simulated measurements from Year 5 of the Dark Energy Survey, we test their robustness to various analysis cases. Our proposed methods produce improved maps and power spectra when compared to current methods, while requiring almost no user tuning.
We end with recommendations for systematics mitigation in future surveys, and note that the methods presented are generally applicable beyond the galaxy distribution to any field with spatial systematics.

\end{abstract}
\maketitle
\section{Introduction}\label{sec:intro}

Over the past 40 years,
cosmological surveys have produced increasingly detailed maps of the large-scale structure (LSS) in the Universe \citep{Lick,deLapparent:1985umo,APM,EDSGC,2dF,6dF,Wiggles,SDSS,BOSS,HSC,kids,Abbott:2017wau}. These observations have proven crucial for testing our understanding of gravity and cosmological structure formation, and helped to constrain cosmological parameters to
the percent level \citep[e.g.][]{Anderson:2013zyy,Alam:2016hwk,Abbott:2017wau}.
Recent observations from DES have for the first time imposed strong constraints on dark energy using an LSS survey \text{alone}, independently of the cosmic microwave background \citep{Abbott:2018wzc}.
Upcoming ground-based missions like DESI \citep{DESI}, and the Rubin Observatory's LSST  \citep{collaboration2012large}, along with space-based missions like SPHEREx \citep{spherex}, Euclid  \citep{Amendola_2018}, and RST (formerly WFIRST)  \citep{spergel2013widefield} will truly herald the age of precision cosmology, mapping up to $\sim$20 billion galaxies across the sky and bringing unprecedented precision to measurements of the dark energy equation of state and modified gravity. 
Such statistical precision makes the control of systematic errors in these datasets of paramount importance to avoid biasing cosmological analyses.

Cosmological information is extracted from LSS observations in multiple ways. The most common approach is to calculate the two-point correlation function or its Fourier counterpart, the power spectrum, to characterize the spatial distribution of galaxies (galaxy clustering) or their shapes (weak lensing). To date, these have been used in cosmological analyses to great success \cite{Hauser_Peebles_1973,Peebles_Hauser_1974,Davis_Peebles,Saunders_IRAS,Fisher_1993,Fisher_1994,Peacock_Nicholson,Feld_Kai_Peac,Baugh_Efstathiou,Baugh_1996,Miller_Batuski,Dodelson_Gaztanaga,Huterer_Knox,Eisenstein_Zaldarriaga,Peacock_2dF,Dodelson_SDSS,Connolly_SDSS,Percival_2dF_Pk,Tegmark_SDSS_Pk,Blake,Elvin-Poole:2017xsf}.
The two-point function contains all available information when the field it characterizes is Gaussian, but nonlinear gravitational collapse induces non-Gaussianity at late times and small scales. Therefore there is considerable cosmological information that is inaccessible to the two-point function. 
This has led to growing interest in using complementary statistical representations of LSS observations, such as higher order N-point functions \citep{Peebles_Groth,Feldman,Scoccimarro-IRAS,Cooray_Hu_bisp,Sefusatti_bisp,Feldman,Scoccimarro-IRAS,Verde_2dF,Marin-wiggles,Gil-Marin,Slepian_Eisenstein_3pt_nlogn,Slepian_Eisenstein_3pt_SDSS}, statistics of peaks \citep{Jain_peaks,Marian_peaks,Liu_peaks} and voids \cite{White_1979,Fry_1986,Biswas,Bos,Pisani,Nadathur,Leclercq}, density-split statistics \cite{Friedrich_2018}, marked power spectra \cite{Sheth_marked,White_2016,White_Padm_marked,Philcox_marked}, Minkowski functionals \citep{Gott_mink,Kratochvil_mink,Hikage_mink,Munshi_mink,Petri_mink,Mawdsley_mink}, wavelet transforms \citep{allys2020_wph, cheng2020_scattering} and more. Further information is also gleaned from the baryon acoustic oscillations and redshift-space distortions. These methods rely on the accurate mapping of the underlying cosmological fields from which they are derived, so there is increasing need for tools to mitigate systematic contamination at the levels of both the map and the two-point functions.

Here, we consider a very general class of systematics that describes an arbitrary spatial modulation of the observed field. Such generic sources of error are one
of the most serious contaminants in our quest to probe cosmology with future surveys.  For definiteness we focus on the case of galaxy clustering, where the systematic error corresponds to a modulation of the galaxy selection function in redshift or across the observing footprint. \textbf{However, the methods we test and develop in this paper are general enough to apply to any real or complex field for which there exist maps of potential contaminants} (e.g.\ shear or Sunyaev-Zeldovich-effect fields). 

Spatially-varying systematics in LSS maps may be caused by an large variety of physical effects. These include
observing conditions and dust extinction (both of which effectively create a
position-dependent ``screen," obscuring background galaxies), bright objects and star-galaxy separation (which can eclipse, change the shape, or be confused for galaxies close to them on the sky), and variations in sensitivity of the detector (which include potentially time- and position-dependent
variations in the focal plane), or imaging pipeline. In all of these cases, failure to fully account for variability in the selection function will result in residual artifacts --- \textit{calibration errors} --- in the final data product and potentially bias results \cite{Huterer:2012zs,Shafer:2014wba,Weaverdyck:2017ovf}.
The presence of calibration errors is evidenced by a number of surveys
 \citep{Vogeley,Scranton,Goto:2012yc, Pullen:2012rd, Ho_2012,
  Ho:2013lda, Agarwal:2013qta, Giannantonio:2013uqa, Agarwal:2013ajb} which 
have shown a significant excess of power at large scales where calibration errors are thought 
to be most prevalent. Recent observations (e.g. from the Dark Energy Survey~\cite{Leistedt:2015kka}) demonstrate however that such contamination is by no means limited to large scales alone. In addition to
adding power, calibration errors induce a multiplicative
effect, coupling different scales and thus affecting all scales in the survey,
including those smaller than the typical size of the calibration systematic
itself \citep{Huterer:2012zs,Shafer:2014wba}.  Much recent work
\citep{Ross:2011cz, Pullen:2012rd, Ho_2012, Agarwal:2013ajb,
  Leistedt:2013gfa, Leistedt:2014wia, delubac_eboss_sys,Rykoff:2015avl, Suchyta:2015qbs, Prakash_2016, Awan:2016zuk, Kalus:2016, Bautista_2018, Kalus:2018qsy, awan2019angular, Rezaie:2019vlz, wagoner_des_sys, kong2020SDSSremoving, ross2020completed} has focused on mitigating these 
systematics in order to probe the underlying cosmology.

The simplest strategy to ameliorate the effects of calibration errors is to simply mask scales or data points suspected having large levels of contamination. More sophisticated strategies include using maps of suspected contaminants --- so-called `templates' --- to correct the observations.
An alternative and complementary approach (e.g.\ \cite{Jasche:2009hz,Jasche:2012kq,Kitaura:2012tu,Wang:2014hia,Jasche:2014vpa,Wang:2016qbz,Modi:2019hnu, Porqueres_2019}) is to forward-model many possible realizations of the cosmic initial conditions. One then evolves these initial conditions in time (while adding realizations of nonlinearities, bias, and observational/instrument systematics), and performs joint inference of cosmology, the initial conditions and late-time ``true" fields, given observations. Yet another  forward-modeling approach involves the injection of false images into observations in order to sample the selection function \cite{Suchyta:2015qbs, kong2020SDSSremoving}.  While such forward approaches are powerful and very general, they also require extensive computational resources and are complicated to implement.
In contrast, using templates to clean contaminated observations and directly infer the underlying fields is straightforward to implement and can be readily incorporated into ongoing or completed analyses. They have been the dominant approach in the community thus far, and so these are the methods we focus on here.

In this paper we revisit and extend state-of-the-art LSS systematics-cleaning strategies.  We interpret them through a  regression framework to highlight commonalities and differences of the methods, as well as some tacit assumptions. In doing so, we show that the common pseudo-$\Cl$ Mode Projection method is equivalent to linear regression. We use this framework to propose straightforward extensions that leverage the extensive body of literature and tools that have been developed for regression analyses. We rigorously test the performance of several existing methods, plus new ones that we propose, on a common set of simulated observations from current and future surveys. 

\begin{figure*}[htbp]
\centering
\includegraphics[width=\textwidth]{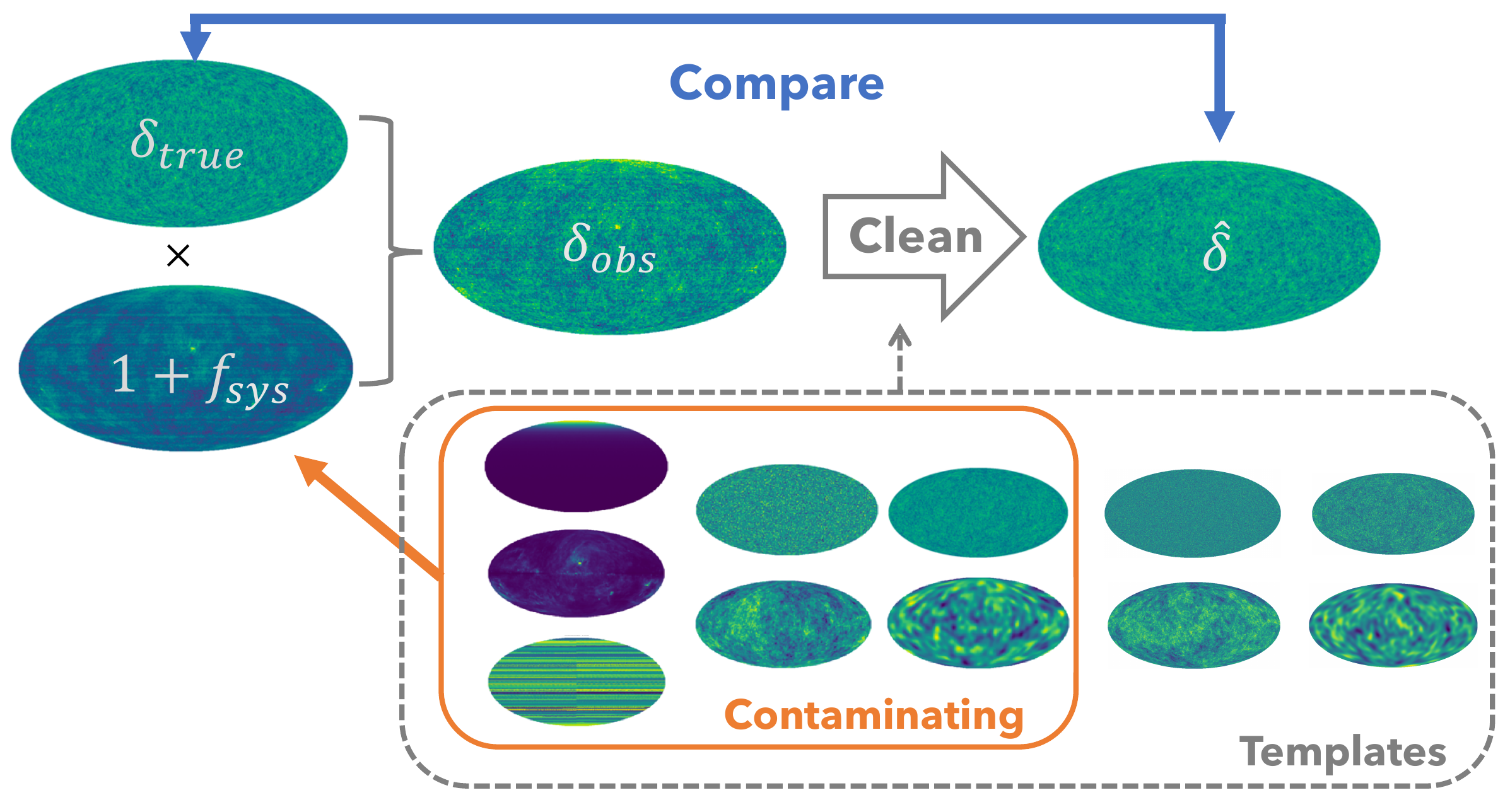}
\caption{Analysis procedure for a single map. A set of templates is generated (dashed box) along with a true overdensity map $\delta_{\rm true}$. A subset of the templates (orange box) contaminate the true overdensity map to generate the observed overdensity field $\delta_{\rm obs}$. We generate an estimated signal map $\hat{\delta}$ using one of the cleaning methods, and compare it to the truth, either at a map-level or power-spectrum-level. This is repeated for many realizations of the signal map and the performance of each cleaning method is assessed.}
\label{fig:flowchart}
\end{figure*}

We study performance using an ensemble of simulated galaxy overdensity maps, such that we can assess both the accuracy and precision of each method. We provide a library of templates and a contaminated overdensity map as input to each cleaning method, which then produces an estimate of the true overdensity map and power spectrum that we assess for accuracy. We repeat the process over a large number of sky realizations and for various configurations of templates to asses the precision and robustness of each method. A schematic outline of this process is shown in Fig.~\ref{fig:flowchart}.

The paper is organized as follows. In Sec.~\ref{sec:calmodel} we describe in detail our general model for contamination, which encompasses a wide range of systematics due to foregrounds or instrument calibration errors. In Sec.~\ref{sec:existmethod} we describe several existing methods for systematics mitigation; in Sec.~\ref{sec:commonframe} we reinterpret the methods through a common framework to facilitate comparison, and in Sec.~\ref{sec:applications} we use this to map several aspects onto well-known techniques in statistics and propose two new mitigation methods. In 
 Sec.~\ref{sec:evalperformance} we describe the fiducial synthetic surveys on which we test the efficacy of the methods that we study. Sec.~\ref{sec:results} shows the results of these performance comparisons, while Sec.~\ref{sec:conclusion} has our conclusions. Several Appendices show important but more technical and detailed aspects of the investigation.

\section{Contamination Model}\label{sec:calmodel}
We first introduce the model for contamination of the observed LSS fields. It is very general, encompassing most known sources of real-world contamination.

Following \citet{Huterer:2012zs}, we model the observed number density map as the product of the true galaxy number density map ($N_\mathrm{true}(\nhat)$) and a direction-dependent screen $(1+\fsys(\nhat))$,
\be
\Nobs(\nhat) = (1 + \fsys(\nhat)) N_\mathrm{true}(\nhat).
\label{eq:model}
\ee
Here, $\fsys(\nhat)$ characterizes the systematic modulation of the true field, such that pixels with $\fsys(\nhat)=0$ are free of contamination. 
Using $N = \bar{N}(1+\delta)$ and defining the ratio of true to observed mean number density as $\gamma = \bar{N}_\mathrm{true}/\bar{N}_\mathrm{obs}$, the observed overdensity can be written as
\be \label{eq:fsysfull}
\delta_\mathrm{obs}(\nhat) = \gamma(\delta(\nhat) + 1)(\fsys(\nhat) + 1) - 1.
\ee
Here $\gamma$ enforces the constraint that $\pixavg{\deltaobs}=0$ across the survey footprint, even though this is not necessarily true for the \textit{true} overdensity field $\delta$. This is due to the fact that we can only access the observed mean number density $\bar{N}_\mathrm{obs}$, which differs from the true mean both because of systematic contamination and because of sample variance from a limited survey footprint (see Sec.~\ref{sec:addmultmethods} for details). 

This model for contamination is similar to the one used in \citep{Huterer:2012zs, Shafer:2014wba, Muir:2016veb, Weaverdyck:2017ovf} to assess the impacts of \textit{residual} calibration errors that remain in the data after cleaning. Here we focus on the methods used to perform such cleaning, and so use the screen model to describe contamination more generally.

We extend the screening formalism by considering that the total systematic modulation is comprised of $\Nsys$ individual systematics, each of which acts as its own screen. Thus we have

\begin{align}
1 + f_{\rm sys} &= \prod_{i=1}^{\Nsys}(1+f_i)\\
                &\simeq 1 + \sum_{i=1}^{\Nsys} f_i + 
                \sum_{j \neq k}^{\Nsys} f_j f_k +
                [\mathrm{higher\, order\, terms}]\nonumber
\end{align}
where we have suppressed $\nhat$ in the notation for convenience both here and in what follows. Note that even if a systematic individually contributes to $\fsys$ linearly, there exist interaction terms with other systematics up to order $\Nsys$. Here and in general, $f_i\equiv f_i(\nhat)$ is a column vector with each element corresponding to a pixel, unless otherwise noted. 

\section{Background: Existing Mitigation Methods}\label{sec:existmethod}
The principal goal of this paper is to compare various proposed systematics mitigation methods. 
The methods that we test are all designed to use maps that trace potential contamination in order to mitigate the impact of systematics, i.e. they assume that the systematic $f_i(\nhat)$ is a function of some tracer $t_i(\nhat)$. We refer to these tracer maps as \textit{templates}, and examples include maps of stellar density, extinction, or summary statistics of observing conditions (e.g. mean g-band seeing) in each region of the sky throughout the duration of the survey (see \citet{Leistedt:2015kka} for a detailed description of the process for creating templates from multi-epoch observational data for the Dark Energy Survey). Sources of error for which we have no templates (e.g. shot noise) are implicitly subsumed into the overdensity field.

We will investigate how effectiveness depends on analysis choices and suggest improvements where possible. We start with three principal methods that have been applied in the literature: the Dark Energy Survey Year 1 method (henceforth DES-Y1), the Template Subtraction method (TS), and the Mode Projection method (MP). While at face value the algorithms associated with these methods seem quite different, we demonstrate that they can be translated into a common mathematical framework of linear regression. Doing so allows us to distill commonalities and differences between the methods, as well to identify simplifications and extensions to them. We include three additional methods based on these insights.

For all the methods, we will work with maps that are divided into pixels in $\texttt{HEALPix}$\footnote{\url{http://healpix.sourceforge.net}}\cite{Gorski:2004by} format, which summarize the mean galaxy overdensity or template values within each pixel (see Sec.~\ref{sec:evalperformance} for details). Furthermore, while we work in the context of cleaning galaxy overdensity fields, the methods are applicable more generally to corrections of any field for which we have templates of potential contamination, and so we denote the true signal more generally as $s$ and the observed field as $\dobs$. In our application, these correspond to the true and observed galaxy overdensity fields, $\delta_{\rm true}$ and $\delta_{\rm obs}$. In the sections that follow, we use $\hat x$ to denote an estimate of $x$, and $\tilde C_\ell^{xx}$ to indicate a realization-specific measurement of the power spectrum, as compared to its theoretical mean $C_\ell^{xx}$.

\subsection{Dark Energy Survey Y1 Method}\label{sec:Y1}

\begin{figure}[htbp]
\includegraphics[width=\linewidth]{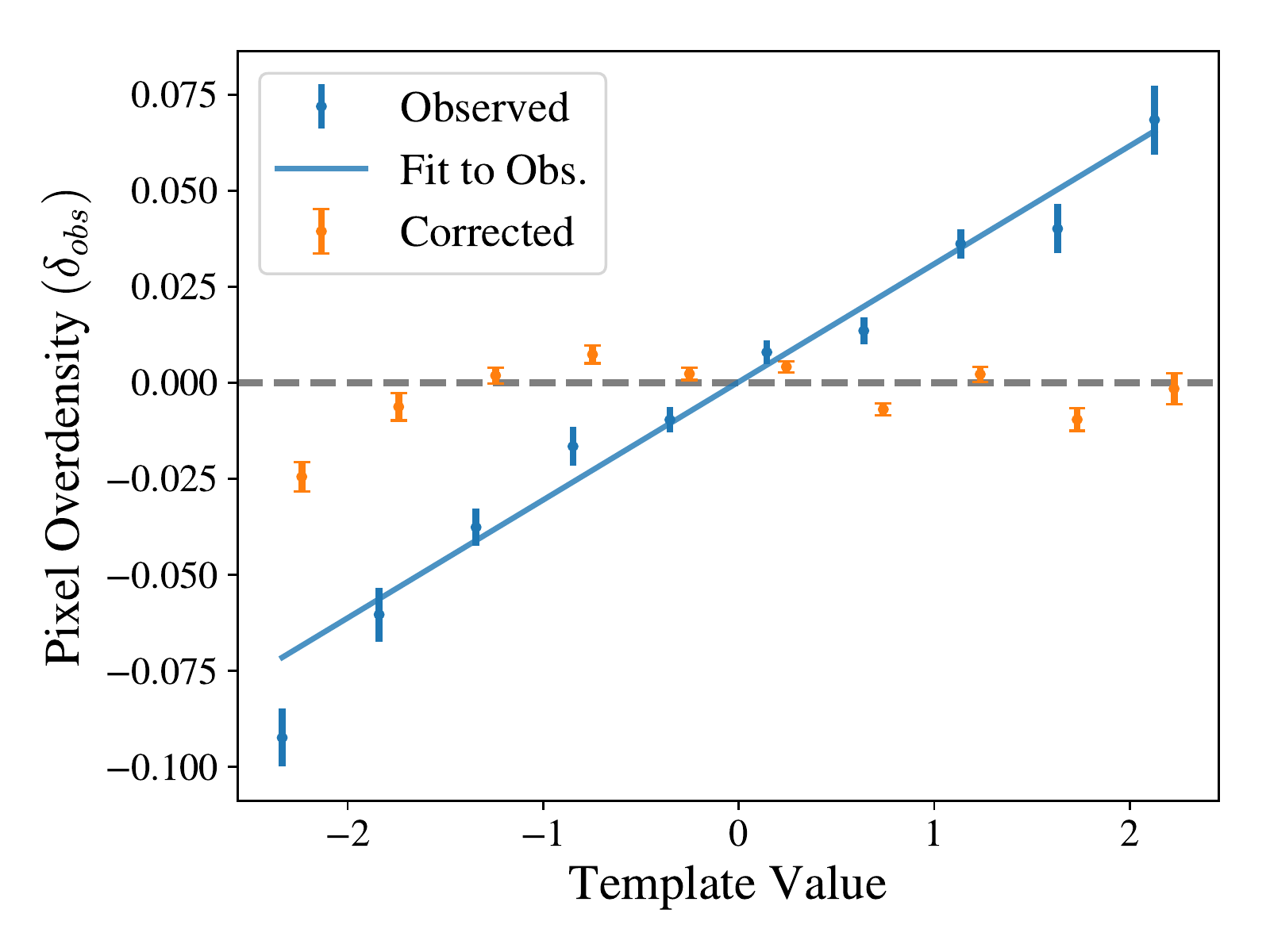}
\caption{Illustration of the DES-Y1 cleaning method, showing the total observed pixel overdensity ($\deltaobs$) as a function of a \textit{template's} pixel overdensity, in ten evenly-spaced bins. Given the estimated covariance matrix (diagonals shown by blue error bars), the best-fit trend (blue line) can be calculated and used to reweight the observed map, producing a corrected map whose dependence on the template is removed (orange points, with corresponding standard errors on the pixel means). 
The process is then iterated for other templates until a satisfactory threshold is reached; see text for details.}
\label{fig:ep1dfit}
\end{figure}

The method used to derive galaxy weights for the Year-1 DES release is one of the more sophisticated mitigation methods applied to date.  It is described in detail in \citet{Elvin-Poole:2017xsf}, but we review its main features here. Hereafter referred to as `DES-Y1,'  it builds on the method first proposed as the `Weights' method in \citet{Ross:2011cz}, wherein 1-dimensional relationships between observed galaxy densities and systematic templates (there called `survey property maps') are removed by iteratively applying multiplicative factors (`weights') to galaxies. Fig.~\ref{fig:ep1dfit} shows one example of how the observed overdensity varies with a template. Multiplicative weights are applied to galaxies to de-trend the data, shifting the blue line to lie atop the dashed line. This method is explicitly a regression method, with versions employing linear fits \cite{Ross:2016gvb, Laurent:2017gze, Ata:2017dya}, splines \citep{Hernandez-Monteagudo:2013vwa} or 
  higher-order polynomials \citep{Nicola:2016eua} as fitting functions for the 1D relationships. 

Here we describe the version we adopt, which closely follows the implementation in \citet{Elvin-Poole:2017xsf} used on the DES-Y1 data. For each template $t_i$, we group pixels into 10 evenly-spaced bins based on their template values, and independent of location on the sky (e.g. all pixels with a mean $i$-band seeing value within 10\% of the max would be grouped). 
We then find the mean galaxy overdensity over the pixels in each bin\footnote{In \citet{Elvin-Poole:2017xsf}, extreme regions are removed by eye: each template is inspected and bins that exhibit an average fluctuation in number density of $>20\%$ are masked, as are regions where visual inspection suggests a deviation from non-monotonic behavior (see their Fig.~3). We neglect this step, as it is difficult to automate robustly and in our tests we found that it did not alter our results.}. A  $10\times10$ covariance matrix of these bin means is estimated by performing the same bin-averaging process on a set of 400 uncontaminated mock maps, generated with a fiducial power spectrum for the overdensity field (we assume the true overdensity power spectrum to generate these mocks).

Next, we use $\texttt{scipy.optimize}$ and the estimated covariance to find the parameters $\{m_i, b_i\}$ of the best-fit line\footnote{\citet{Elvin-Poole:2017xsf} also use linear fits for almost all templates, with only a couple exceptions. As noted in Sec.~\ref{sec:addmultmethods}, even if a template is thought to contaminate non-linearly, the relationship can usually be made linear through an appropriate transformation of the template.} of the binned overdensity to each binned template $i$:
\be
\frac{\langle\Nobs\rangle_{j}}{\bar{N}_{\rm obs}} = m_i \langle t_i\rangle_{j} + b_i
\label{eq:Y1_slopes}
\ee
where $\langle \cdot \rangle_j$ indicates the average pixel value in bin $j$ of the given template. See Fig.~\ref{fig:ep1dfit} (blue points and trend) for an illustration.

The template with the most significant fit is used to reweight the number density in each pixel as $\Nobs'(\nhat) = \Nobs(\nhat)/(\hat{m}_i t(\nhat) + \hat{b}_i)$,  where the significance metric is defined below. Having removed the effect of the dominant systematic, the whole process is repeated: for each template, the pixels are assigned to bins and averaged, the new best-fit parameters are computed from Eq.~(\ref{eq:Y1_slopes}), and the trend from the most significant template is removed from the data. The process stops when all templates are below a predefined significance threshold. 

In general, the more contamination from template $i$, the stronger relationship the relationship with the observed galaxy density. However, some level of correlation is expected just by chance, and this depends on the spatial clustering of each template. The DES-Y1 method addresses this in two ways: (1) by using a different covariance matrix for the observed overdensity for each template as described above, and (2) by having a template-specific significance threshold, calibrated on mocks. Specifically, the significance statistic used is $\Delta\chi_i^2/[\Delta\chi_i^2]_{68}$, where $\Delta\chi_i^2$ is the improvement in $\chi^2$ for the binned fit on template $i$, compared to a null hypothesis of $m_i=b_i=0$. It is normalized to the $68^{\rm th}$ percentile of the same quantity measured on uncontaminated signal mocks ($[\Delta\chi_i^2]_{68}$). 
We use the stopping criterion $\Delta\chi_i^2/[\Delta\chi_i^2]_{68} < \Delta\chi^2_{\rm threshold} = 2$, but find that our results change little when changing this threshold between 1 and 4 (see App.~\ref{app:Y1}).

There are a number of required parameter choices in the DES-Y1-type method. These include the  criterion for selecting the most significant template,\footnote{E.g. one could consider an $R^2$ statistic, the commonly-used $F$-statistic, Akaike or Bayesian information criteria, etc.} the significance threshold that determines when to stop weighting, the prior power spectrum for generating mocks, and choices associated with binning (e.g. number of bins, equally-spaced vs. equally-filled, etc). Here we use the fiducial choices from \citet{Elvin-Poole:2017xsf}, and investigate some of the effects of these choices in App.~\ref{app:Y1}.

\subsection{Template Subtraction}
The Template Subtraction method uses the cross-power of templates with the observed  sky to estimate contamination of each template at each angular scale. Contamination is subtracted directly from the two-point clustering statistics. The method was proposed in \citet{Ho_2012} and \citet{Ross:2011cz} where it was called the ``cross-correlation" technique, and we review it here. 

Template Subtraction assumes the observed overdensity $\dobs$ is a linear combination of the true galaxy overdensity $s$ and individual template overdensities $t_i$:
\be
d_{\rm obs} = s + \sum_{i=1}^{N_{\rm tpl}} \alpha_i t_i.
\label{eq:linaddmodel}
\ee
Any systematics or noise not accounted for by templates are subsumed into the signal $s$. In \citet{Ho_2012}, $\dobs$ and $t_i$ are taken to be in multipole space, such that $\langle s s\rangle \rightarrow \langle s_{\ell m} s_{\ell m}\rangle = \Clss$ (where $\langle \cdot \rangle$ is the ensemble average over many sky realizations), and $\alpha \rightarrow \alpha_\ell$ is a function of $\ell$. The companion paper of \citet{Ross:2011cz} works in configuration space, so in their version of Template Subtraction, the data vectors are in pixel-space. 

We will work in harmonic space and so follow \citet{Ho_2012}, but we will keep the notation general until dealing with the two-point functions where we will explicitly work with power spectra. The treatment for configuration space is largely identical. To apply the method, one would simply substitute the correlation function for the power spectrum $\Cl^{ij} \rightarrow w^{ij}(\theta)$ and $\alpha_\ell \rightarrow \alpha(\theta)$. See e.g. \citet{Crocce:2015xpb} for an application of template subtraction to the correlation function.

If we consider just a single contaminant for simplicity ($\Ntpl=1$), and assume that it is uncorrelated with the underlying galaxy field, then from Eq.~(\ref{eq:linaddmodel}) the two point function of the observed field is 
\be
\langle \dobs \dobs \rangle = \langle s s \rangle + \alpha^2 \langle t t \rangle.
\ee

Then on average, 
\begin{align}
\langle t \dobs\rangle / \langle t t \rangle = \Cl^{td}/ \Cl^{tt} = \alpha_\ell
\end{align}
and the contamination at each multipole can be estimated as
\begin{align}
\hat \alpha_\ell &= \tilde \Cl^{td}/ \Cl^{tt}
\label{eq:ts1tpl}
\end{align}
where the tilde in $\tilde{C}_\ell$ indicates the power spectrum that is measured from the observed sky realization, and $\tilde{C}^{tt}_\ell = \Cltt$ since we take templates to be fixed.

An estimate of the power spectrum can then be found to be 
\be \label{eq:tsclss1}
\hat{C}_\ell^{ss} = \left(\tilde{C}_\ell^{dd} - \hat{\alpha}_\ell^2 \Cltt\right)
\left(1 - \frac{1}{2\ell+1}\right)^{-1}.
\ee 
Here $[1 - 1/(2\ell+1)]^{-1} = [(2\ell+1)/(2\ell)]$ 
is a factor found by \citet{Elsner:2015aga} that is needed to debias the estimator.\footnote{In the case of the correlation function, the bias cannot be written in a signal-independent fashion, and so requires a prior signal power spectrum or simulations to estimate.} 
The bias arises because the process is too aggressive --- any chance correlation between template and the true signal is also removed, resulting in an underestimate of the true clustering power. 

The Template Subtraction method is easily generalized to multiple templates by extending
the dimensionality of terms as 
\begin{align*}
    \alpha_\ell~{\rm (scalar)} &\rightarrow \balpha_\ell ~(\Ntpl)\\
    \Cltt ~{\rm (scalar)} &\rightarrow {\bf \ClTT} ~(\Ntpl \times \Ntpl)
\end{align*}
and Eqs.~(\ref{eq:ts1tpl}) and (\ref{eq:tsclss1}) become 
\be\label{eq:tsalphamult}
\mathbf{\hat{\balpha}}_{\ell} = [{\bf \tilde C_\ell^{TT}}]^{-1} [{\bf \tilde C_\ell^{Td}}],
\ee
\be \label{eq:tsclssmult}
\hat{C}_\ell^{ss} = \left(\tilde{C}_\ell^{dd} - \hat{\balpha}^\dag\ClTT\hat{\balpha}\right)
\left(\frac{2\ell+1}{2\ell+1 - \Ntpl}\right).
\ee
For the cut-sky equations, we refer the reader to Ref.~\cite{Elsner:2015aga}. 

While previous work on Template Subtraction has focused on the cleaned power spectrum, an estimate of cleaned overdensity field itself is also of interest for cosmological study, as it contains more information than just its power spectrum. 

A map estimate from the Template Subtraction method can be produced as
\be
\shat^{TS}(\nhat) = \sum_{\ell=1}^\infty \sum_{m=-\ell}^\ell \shat_{\ell m}^{TS}\, Y_{\ell m}(\nhat),
\ee
where the harmonic coefficients of the map are given by
\be
\shat_{\ell m}^{TS} = (\dobs)_{\ell m} - \sum_{i=1}^{\Ntpl} (t_{\ell m})_i (\hat\alpha_\ell)_i \label{eq:sesttsalpha}
\ee
and the (biased) power spectrum of the cleaned map is equivalent to the first factor in Eq.~(\ref{eq:tsclssmult}). 

\subsection{Mode Projection}
Mode Projection (also often called Mode \textit{De}projection \cite{Kalus:2016, Percival:2018vsn, Kalus:2018qsy, Alonso:2018jzx, nicola2019tomographic}) assumes the same contamination model as Template Subtraction, given by Eq.~(\ref{eq:linaddmodel}). The original formulation \cite{Rybicki:1992jz, Leistedt:2013gfa} cleans the map-level systematics by assigning infinite variance to contaminating templates. This procedure desensitizes the power spectrum estimate to the templates and is equivalent to marginalizing over the contamination amplitude of each template \cite{Leistedt:2013gfa}. 

In particular, it updates the map-level covariance matrix ${\bf C}$ as follows
\begin{equation}
\begin{aligned}
{\bf C'} &=  \left[{\bf C} + \sum_k^{\Ntpl} \lim_{\beta \to \infty}(\beta_k t_k t_k^\dag)\right]\\
        &= \lim_{\beta \to \infty}\left[{\bf C} + \beta TT^\dag\right]
\end{aligned}
\end{equation}
where $t_k$ are the individual template maps, which can represent either real spin-0 or complex spin-2 fields \cite{Alonso:2018jzx}, and which can be assembled into 
a matrix $T$, with $t_k$ as the $k^{\rm th}$ column. In previous works with Mode Projection, the maps have been represented in pixel space, but in principle the operations can also be performed in harmonic space, e.g. representing a spin-0 field by its complex harmonic coefficients. 
For clarity and continuity, we will assume the maps are $\Npix$-length vectors in what follows, as opposed to their multipole transforms. There are some benefits to performing Mode Projection in harmonic space, however, which we explore in Sec.~\ref{sec:commonframe}.

The main challenge with the original formulation of Mode Projection is that it requires the construction and inversion of a covariance matrix for the whole map, which is often intractable. To remedy this, \citet{Elsner:2016bvs} extended Mode Projection to the popular (albeit sub-optimal) pseudo-$\Cl$ estimator. In practice, this is achieved by computing the pseudo-$\Cl$s of the overdensity field after first applying a filter ${\bf F}$, where
\begin{equation}
\begin{aligned}
    {\bf F} &= \lim_{\beta \to \infty} \left(I + \beta T T^\dag \right)^{-1} \\
            &= I - T (T^\dag T)^{-1} T^\dag, 
\end{aligned}
\end{equation}
where the second expression follows from the Sherman-Morrison-Woodbury formula. It is easy to see that $T(T^\dag T)^{-1}T^\dag$ is a projection matrix, projecting an $\Npix$-dimensional map onto a $\Ntpl$-dimensional subspace. The filter thus removes any components of the observed map within the subspace spanned by the templates (hence the alternate name of Mode \textit{De}projection).

Taking the case of a single template map $t$ for simplicity, 
${\bf F}$ then takes the form $(I - (t t^\dag)/(t^\dag t))$, resulting in a filtered overdensity map
\begin{align}
    \shat &= {\bf F}\,\dobs \\
          &= \left[I - t (t^\dag t)^{-1} t^\dag \right]\dobs \label{eq:sestmp}\\
            &= \dobs - t \hat \alpha_{\rm mp} \label{eq:sestmpalpha}
\end{align}
where
\begin{equation}
\hat\alpha_{\rm mp} = (t^\dag \dobs)/(t^\dag t) = \tilvar{td}/\var{tt}, \label{eq:alphampest}
\end{equation}
and $\tilvar{td}$ is a measure of the covariance of maps $t$ and $d$. Note that this is very similar to the Template Subtraction estimate in Eq.~(\ref{eq:ts1tpl}), but here the covariances are taken over the whole footprint, rather than for a single mode $\ell$. We can make the connection even more explicit by noting that in the full-sky case,
\be 
\tilvar{td} = \frac{1}{4\pi}\sum_{\ell=0}^\infty (2\ell + 1)\tilde{\Cl}^{td}
\ee
While \citet{Elsner:2016bvs} introduce this filtered map only as a means to compute the power spectrum, it can be used on its own as an estimate for the cleaned overdensity field. However, as with Template Subtraction, the power spectrum of this cleaned \textit{map} is a biased estimate of the true power spectrum, as some of the signal is removed in the cleaning process: 

\begin{align}
      \langle \Cl^{\shat \shat}\rangle &= \langle (\dobs - t \hat \alpha_{\rm mp})^\dag (\dobs - t \hat \alpha_{\rm mp})\rangle \\
      &= \Clss - \frac{\Cltt}{4\pi\left(\sigma^2_{tt}\right)^2}
      \left(2\Clss\var{tt} -  \frac{1}{4\pi}\sum_{\ell'} (2 \ell' +1)C_{\ell'}^{ss}C_{\ell'}^{tt}\right).
      \label{eq:clsrawsrawmp}
\end{align}

In the full sky case, the power spectrum estimate can be debiased analytically \cite{Elsner:2016bvs}:

\begin{align}
    \hat{C}_\ell^{ss} &= \sum_{\ell'} \left[(I + B)^{-1}\right]_{\ell \ell'}C_{\ell'}^{\shat \shat},
    \label{eq:clestmp}
\end{align}
where
\be
B_{\ell \ell'} = \frac{\Cltt}{4\pi\left(\var{tt}\right)^2}\left(-2\vartt\delta_{\ell \ell'} + \frac{2 \ell' +1}{4\pi}C_{\ell'}^{tt}\right)
\ee
and $\delta_{\ell \ell'}$ is the Kronecker delta.
In the presence of a mask, one can debias via iteration or assuming a prior power spectrum \cite{Elsner:2016bvs}. As we work in the full-sky case, we debias analytically via Eq.~(\ref{eq:clestmp}), though we do not expect an iterative or prior-based debiasing to significantly alter our conclusions.

The procedure outlined above easily generalizes to multiple maps by extending the dimensionality of the terms:
\begin{eqnarray}
    \nonumber
    \alpha~{\rm (scalar)} &\rightarrow& \alpha ~(\Ntpl),\\
    t ~{(\Npix}) &\rightarrow &{\bf T} ~(\Npix \times \Ntpl),\\
    \nonumber
    \var{td} ~{\rm (scalar)} &\rightarrow& {\bf \var{Td}} ~(\Ntpl),\\
    \nonumber
    \Cltt ~{\rm (scalar)} &\rightarrow& {\bf \ClTT} ~(\Ntpl \times \Ntpl),\\
    \nonumber
    \var{tt} ~{\rm (scalar)} &\rightarrow& {\bf \var{TT}} ~(\Ntpl \times \Ntpl).
\end{eqnarray}

Hereafter, we will use `Mode Projection' to refer to the pseudo-$\Cl$ mode projection method described above, due to the popularity of the pseudo-$\Cl$ power spectrum estimator and the adoption of this version into \texttt{NaMaster}\footnote{\url{https://github.com/LSSTDESC/NaMaster}}\cite{Alonso:2018jzx}, in anticipation of LSST. We again refer the reader to \citet{Elsner:2016bvs} for the modifications necessary to account for the mask, and specifically to their Eq.~(21) for the multi-template version of the debiasing matrix, which we use to correct for all Mode Projection power spectrum estimates (see Ref.~\cite{Alonso:2018jzx} for the equivalent formulae for spin-2 fields).

\section{Placing into a Common Mathematical Framework}\label{sec:commonframe}
To facilitate a comparison of the methods, it is useful to place them into a common mathematical framework. In this section, we show how all three methods presented so far can be interpreted through a regression analysis lens, and in doing so help identify different assumptions within each method and possible avenues for improvement. Moreover, we can leverage the powerful suite of tools that have already been developed and tested for regression to the task of systematics removal, facilitating and accelerating the process. 

\subsection{Connections to Regression}\label{sec:regression}
\begin{figure*}[htbp]
\includegraphics[width=\linewidth]{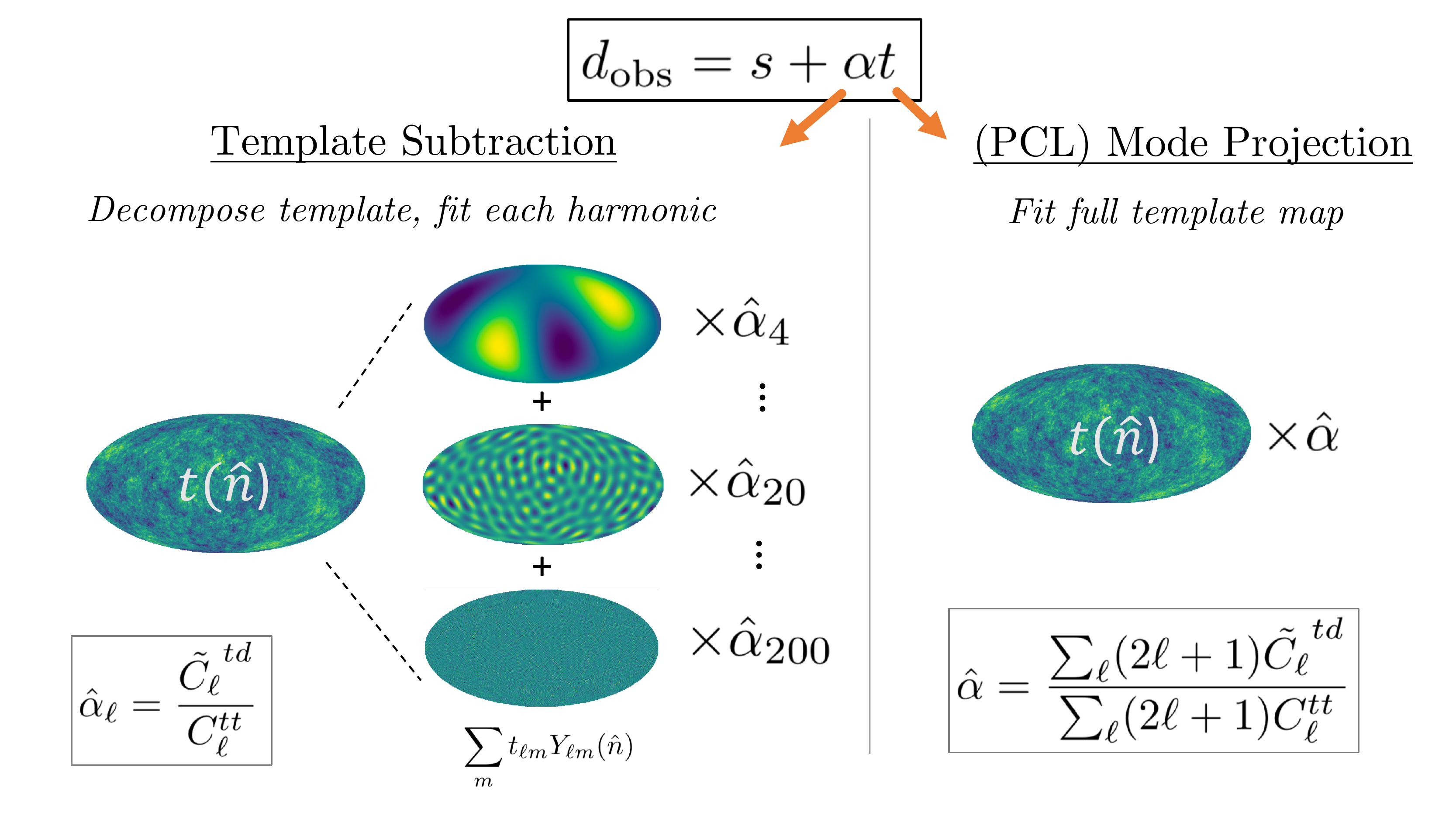}
\caption{Schematic illustration of the difference between the Template Subtraction and (pseudo-$C_\ell$) Mode Projection methods. Template Subtraction allows templates to have different levels of contamination at each scale. This is analogous to performing Mode Projection, but first decomposing each template map into a series of derived templates, each corresponding to a different harmonic $\ell$. See Sec.~\ref{sec:regression} for details.}
\label{fig:mptsdiffdiag}
\end{figure*}

We have purposefully formulated the methods (e.g. Eqs.~(\ref{eq:alphampest} and (\ref{eq:ts1tpl}))in a manner designed to make the connections between Mode Projection and Template Subtraction apparent. 
Template Subtraction is equivalent to running the Mode Projection algorithm, but with each original template ($t_i(\nhat)$) decomposed into a set of independent templates ($t^i_\ell(\nhat)$), where 
\be
t^i_\ell(\nhat) = \sum_{m=-\ell}^\ell \, t^i_{\ell m}Y_{\ell m}(\nhat).
\ee
Fig.~\ref{fig:mptsdiffdiag} shows this schematically.
In other words, (pseudo-$C_\ell$) Mode Projection can be considered a special case of Template Subtraction, where the contamination is assumed to be independent of scale \textit{and} the full template map is used to estimate such contamination.
It has been pointed out before in the context of 3D clustering estimates that Template Subtraction and Mode Projection can be related if they use equivalent templates \cite{Kalus:2016}.

Casting the two methods into this form allows us to make the connection to standard linear regression wherein a measured response $\mathbf{y}$ is assumed to be a linear combination of predictors given by the $\bm{\alpha}$ and a noise term $\bm{\epsilon}$: 
\be
\mathbf{y} =  \mathbf{X}\mathbf{\alpha} + \mathbf{\epsilon}.\label{eq:linmodelgen}
\ee
$\mathbf{X}$ is a $n\times p$ matrix, where $p$ is the number of predictors (potentially including a column of ones --- the intercept term), $\bm{\alpha}$ a vector of length $p$, and $\mathbf{y}$ and $\bm{\epsilon}$ vectors of length $n$.

Perhaps the most common regression method, Ordinary Least Squares (OLS), finds the vector $\hat\alpha$ that minimizes the squared residuals:
\begin{align}
\hat{\bm{\alpha}} &= {\rm argmin}_\alpha ||\mathbf{y - X}\bm{\alpha}||^2 \label{eq:alphaargmin}\\
                &= (\mathbf{X^\dag} \mathbf{X})^{-1}\mathbf{X^\dag} \mathbf{y},
\end{align}
where the second expression follows if $X$ is full column-rank (i.e. the number of observations exceeds the degrees of freedom from the predictors).
This is equivalent to the maximum likelihood solution if one assumes the noise of each element, $\epsilon_i$, is independent and identically Gaussian distributed, 
\be
P(\bm{y}|\bm{X}\alpha)\sim \mathcal{N}(0,I\sigma_\epsilon).\label{eq:gausslike}
\ee
such that the log-likelihood goes as $\mathcal{L} \propto |\mathbf{y - X}\bm{\alpha}|^2$.
Even if the assumption of Gaussianity is violated, by the Gauss-Markov theorem Eq.~(\ref{eq:alphaargmin}) still corresponds to the unbiased estimator with minimum variance if the errors $\bm{\epsilon}$ are uncorrelated and have equal variance.

We can write Eq.~(\ref{eq:linmodelgen}) in terms of the OLS estimates as 
 \begin{equation}
    \mathbf{y} = \mathbf{X}\mathbf{\hat\alpha} + 
    \mathbf{\hat \epsilon} 
    = \mathbf{X}(\mathbf{X^\dag} \mathbf{X})^{-1}\mathbf{X^\dag} \mathbf{y} + \mathbf{\hat\epsilon} 
    \label{eq:yxalphaeps}
\end{equation}
where the residuals are defined as
\begin{equation}
          \mathbf{\hat\epsilon}  = \mathbf{y} - 
          \mathbf{X}\mathbf{\hat\alpha}. \label{eq:epshat}
\end{equation}
The quantities of interest in the typical regression problem are the coefficients $\bm{\alpha}$ or the predicted response $\hat{\mathbf{y}}=\mathbf{X}\hat{\bm{\alpha}}$, with the goal of understanding the influence of predictors or to predict future observations, and hence the residuals are largely used to assess whether the basic OLS assumptions hold. However comparing Eq.~(\ref{eq:epshat}) to Eqs.~(\ref{eq:sesttsalpha}) and (\ref{eq:sestmpalpha}), we see that \textbf{both Mode Projection and Template Subtraction can be interpreted as OLS regression methods where the observed overdensity signal is regressed onto the templates, and the reconstructed overdensity signal $\shat$ and power spectrum $C_{\ell}^{\shat \shat}$ correspond to the map and power spectrum of the residuals $\bm{\hat\epsilon}$}. 

Mode Projection uses the full map footprint, with each pixel corresponding to a single observation,
for a total of $\Ntpl$ fit coefficients. 
In contrast, Template Subtraction can be interpreted as performing multiple OLS regressions in parallel on smaller subspaces --- one at each multipole in our case --- for a total of $N_\ell \times \Ntpl$ fit coefficients (see Fig.~\ref{fig:mptsdiffdiag}).

We can write the Template Subtraction amplitudes computed by Eq.~(\ref{eq:tsalphamult}) in OLS form as 
\be
\hat\alpha_\ell = (\bfTell^{\,\dag} \bfTell)^{-1} \bfTell^{\,\dag} {d}_\ell,
\ee
where $\bfTell$ is a $(2\ell+1) \times \Ntpl$ matrix, with each column corresponding to a template, consisting of all the harmonic coefficients for a fixed $\ell$: 

\be
\label{eq:Tlm}
\bfTell = \left(\begin{array}{cccc}
  t^1_{\ell,-\ell} & t^2_{\ell,-\ell} &\cdots &t^{\Ntpl}_{\ell,-\ell}\\
  t^1_{\ell,-\ell+1} & t^2_{\ell,-\ell+1} &\cdots &t^{\Ntpl}_{\ell,-\ell+1}\\
  \vdots&\vdots &\ddots &\vdots\\
  t^1_{\ell,\ell} & t^2_{\ell,\ell} &\cdots &t^{\Ntpl}_{\ell,\ell}\\
\end{array}\right). 
\ee

In cases where the multipoles (or angular scales) are binned, the number of fit coefficients is reduced to $N_{\rm bins}\times\Ntpl$, which reduces the variance of the contamination estimate. Indeed, Mode Projection corresponds to a limiting case, where the modes of each template are averaged with equal weight before fitting. However in principle one could apply weights differently across scales, and as we will show, this can produce improved coefficient estimates. Alternatively, one could fit individual modes as in Template Subtraction but combine at the coefficient level --- potentially useful if certain scales are of particular interest for a given analysis.\footnote{In their real-space analysis of SDSS galaxies, \citet{Ross:2011cz} seem to implement a version of this. They use Template Subtraction to produce fit coefficients for a large number of scales and templates, but ultimately select one coefficient to apply to all scales for each template However it is unclear how they compute the single summary coefficient.}

An immediate consequence of the OLS interpretation of these methods is in making explicit the assumptions that Mode Projection and Template Subtraction are making about the underlying density field --- they are exactly the ``OLS" assumptions for the error term $\epsilon$ in the regression model: independent, Gaussian and of equal variance, in whatever basis the map is represented. 
These assumptions hold well for Template Subtraction, which performs a separate regression at each multipole $\ell$. In this case, the assumed OLS ``noise" terms are the set of harmonic coefficients of the map ($s_{\ell m}$) at that multipole, which have 
${\rm Cov}[s_{\ell m_1},s_{\ell m_2}] = \Clss \delta_{m_1 m_2}$.
For Mode Projection, these assumptions are violated, as the covariance matrix between overdensity pixels is not diagonal, ${\rm Cov}[s(\nhat_i),s(\nhat_j)] \neq \sigvar\delta_{ij}$. 

Since the primary contribution to the ``noise" of the OLS fit is the clustering signal itself, we can diagonalize it by performing Mode Projection in multipole space, with the maps $d$, $s$ and $t_i$ becoming complex column vectors comprised of the map spherical harmonic coefficients. The noise of the observed overdensity $d_{\ell m}$ is then ${\rm Cov}[s_{\ell_1 m_1},s_{\ell_2 m_2}] = \Clss \delta_{\ell_1 \ell_2}\delta_{m_1 m_2}$. While diagonal, this varies strongly with $\ell$ and therefore violates the assumption of equal variance, a property known as `heteroskedasticity' in the statistics literature.

However once the noise is diagonal, we can improve the Mode Projection estimate of $\hat{\bm{\alpha}}$ by weighting the observed data and template modes by (a prior-inferred) $1/\sqrt{\Clss}$:
\be 
\hat{\alpha} = \frac{\sum_{\ell=0}^\infty (2\ell + 1)\tilde{\Cl}^{td}/\Clss}{\sum_{\ell=0}^\infty (2\ell + 1)\tilde{\Cl}^{tt}/\Clss}.
\ee
This is equivalent to a weighted least-squares approach and recovers the maximum likelihood estimate of $\bm{\hat{\alpha}}$, eschewing the erroneous assumption of a flat signal power spectrum. This of course only works in the ideal full-sky case, but in principle it should not be difficult to extend to a masked sky, e.g. using a predicted cut-sky $\Clss$ computed using the standard coupling matrix from the mask (e.g. \cite{Hivon:2001jp, Elsner:2016bvs}) along with the cut-sky harmonics of the templates and datavector, or appropriate binning of modes. This can be viewed as a form of `prewhitening' the data, which accounts for the off-diagonal pixel covariance in the likelihood through an appropriate transform. We explore the potential improvement from such prewhitening
in App.~\ref{app:prewhitening}, finding that it improves cleaning, but is subdominant to differences between cleaning methods and higher-order corrections we discuss below.

Finally we note that both the Template Subtraction bias from \citet{Elsner:2015aga}, as well as the pseudo-$C_\ell$ Mode Projection bias from \citet{Elsner:2016bvs} result trivially when interpreting them through the OLS lens, in which the variance of observed residuals is well-known to be biased low:
\be
\langle\hat{\epsilon}^\dag\hat{\epsilon}\rangle = \left(\frac{N_{\rm data} - p}{N_{\rm data}}\right) \epsilon^\dag\epsilon.
\ee
For Template Subtraction, the regression at each harmonic has $N_{\rm data} = 2\ell+1$ and number of predictors $p = \Ntpl$, leading exactly to the debiasing terms for the signal power estimate in Eqs.~(\ref{eq:tsclss1}) and (\ref{eq:tsclssmult}). The debiasing terms for Mode Projection in Eq.~(\ref{eq:clsrawsrawmp}) are more complicated and dependent on the signal and template clustering, but if we take both $\Clss$ and $\Cltt$ to be independent of $\ell$, Eq.~(\ref{eq:clsrawsrawmp}) reduces to 
\be
\langle\hat{s}\hat{s}\rangle = C^{ss} \left(1 - \frac{1}{\sum_{\ell'=0}^{\ell_{\rm max}}(2\ell'+1)}\right),
\ee
where $p=\Ntpl=1$ and $N_{\rm data} = \sum_{\ell'=0}^{\ell_{\rm max}}(2\ell'+1) = (\ell_{\rm max}+1)^2$ is the total number of Fourier modes in the map. This is in keeping with the interpretation of \citet{Elsner:2016bvs}, wherein each template removes one degree of freedom from the number of observed Fourier modes. This interpretation can help to assess the risk of overfitting based on the size of the template library.

By making connections between current methods and linear regression explicit, we not only facilitate their interpretation, but can more easily identify the tacit assumptions within these methods, as well as readily improve upon them, drawing on the large body of research into the statistical properties of various regression approaches.

\subsection{Additive vs. Multiplicative Treatment} \label{sec:addmultmethods}

One fundamental way in which the methods described here differ is whether or not they assume the systematic contamination is purely additive. This amounts to neglecting a multiplicative term that, if unaddressed, can bias cosmological constraints in upcoming surveys \cite{Shafer:2014wba}. Here we show how Mode Projection (or any regression method) and be readily adapted to account for the multiplicative contamination and so lead to improved map and power spectrum estimates.

Using the same notation we have used for the additive methods, the general expression for the observed overdensity  is
\begin{equation}
\begin{aligned}
\label{eq:dobsgamma}
    \dobs &= \gamma (1 + s)(1 + \fsys) - 1 \\[.1cm]
\end{aligned}
\end{equation}
where again we have suppressed the pixel index. The factor  $\gamma=\Nbar_{\rm true}/\pixavg{\Nobs}$ accounts for the so-called integral constraint, wherein the mean \textit{observed} number density is used to compute the overdensity field, rather than the true full-sky mean density (see App.~\ref{app:monopole} for a more in depth look at the impact of this monopole term).

To compare with the additive methods, it is convenient to define a zero-centered systematic as: 
\be
f_i' \equiv \frac{f_i - \bar{f}_i}{1 + \bar{f}_{i}}, \label{eq:fprime}
\ee
and write Eq.~(\ref{eq:dobsgamma}) in an equivalent but zero-centered form:
\be
\dobs = \gamma'(1 + s)\left(1 + \fsys'\right) - 1,
\label{eq:dobsgammaprime}
\ee
with the new prefactor
\be
\gamma' \equiv \gamma (1+\bar{f}_{\rm sys}) = \left(1 + \pixavg{s'\fsys'} + \fic\right)^{-1}
\ee
ensuring that the monopole in $\dobs$ is zero, and having the property that $\langle \gamma' \rangle\approx 1$.\footnote{Here the approximation stems from making the assumption $\langle x^{-1} \rangle \approx \langle x\rangle^{-1}$, which holds very well for the cases we are studying where the mean is taken over a footprint with $\Npix \gtrsim 10^5$ and shot noise is subdominant.} Here $\fic\equiv \pixavg{s}$ characterizes the global overdensity in which the footprint resides, and $s' \equiv s - \fic$ is the deviation from that local overdensity. 

Thus the observed overdensity field contaminated with a generic systematic $f_i$ can be equivalently written as contamination from a \textit{zero-centered} systematic with a rescaled amplitude, $f'_i$.

Expanding Eq.~(\ref{eq:dobsgammaprime}), we have
\begin{align}
\dobs 
    &= s + \gamma'\fsys' + \gamma's\fsys' + (\gamma'-1)(s+1),
    \label{eq:dobsexpanded}
\end{align}
Comparing to additive models like Mode Projection and Template Subtraction, which take
\be
d^{\rm add}_{\rm obs} = s + f'_{\rm sys} = s + \sum_{i=1}^{N_{\rm tpl}} f'_i,
\label{eq:dobs}
\ee
we see that they not only assume $\gamma'=1$ (a vanishing correlation between signal and systematics over the full footprint, as well as no mean local overdensity), but also that $s\fsys'=0$ \textit{for every pixel}, a much stronger assumption.
However, despite these assumptions, additive estimates of the total contamination are unbiased, provided the templates $T$ span the space of the true contamination:
\begin{align}
\langle \hatfsys\rangle = \langle T\hat\balpha\rangle &= \langle T(T^\dag T)T^\dag \dobs\rangle \label{eq:hatfsysadd}\\
&\approx T(T^\dag T)T^\dag\fsys' = \fsys'.
\end{align}
Intuitively this makes sense, since in the ensemble average the multiplicative term $sf'$ will vanish.

From Eq.~(\ref{eq:dobsgammaprime}), we can then make an improved estimate of the signal map as
\begin{align} 
\shat &= \frac{1 + \dobs}{1 + \hatfsys} - 1,\label{eq:debiasedols2}\\
     &=\frac{\dobs - \hatfsys}{1 + \hatfsys} \label{eq:debiasedols22}
\end{align}
where the second form makes clear that this is a simple rescaling of the additive signal estimate, $\dobs - \hatfsys$. Therefore, \textit{in a model with multiplicative contamination, signal estimates from additive methods can be improved by weighting the estimated signal map by $1/(1 + \hatfsys)$}. 

To explicitly close the loop on the aforementioned methods, the DES-Y1 method performs a series of 1-D regressions and iteratively weights the observed overdensity in a manner equivalent to Eq.~(\ref{eq:debiasedols2}) for each template, whereas Mode Projection estimates contamination via a single $\Ntpl$-dimensional regression, with a signal estimate that can be improved via Eq.~(\ref{eq:debiasedols22}).\footnote{
As noted in Sec.~\ref{sec:calmodel} even linear contaminants will have interaction terms up to order $\Ntpl$, such that in principle, for Eq.~(\ref{eq:hatfsysadd}) to fully capture $\fsys$, additional templates up to $t_i t_j t_k ...t_{\Ntpl}$ would need to be included in the template library. A more precise and efficient approach would be to not add any interaction templates, but instead combine the base systematic estimates as
\be
\hat{f}_{\rm sys, alt} = \prod_{i=1}^{\Ntpl} (1 + \hat{f_i}) = \prod_{i=1}^{\Ntpl} (1+\hat{\alpha}_i t_i),
\ee
where recall $t_i$ corresponds to the $i^{\rm th}$ template and $i^{\rm th}$ column of $T$, and $\hat\alpha_i$ the $i^{\rm th}$ element of $\hat\alpha$.
This is closer to the treatment of the DES-Y1 method, wherein weightings for each $\hat{f}_i$ are applied in series and thus cumulatively. 

In practice we find that Eq.~(\ref{eq:hatfsysadd}) is a very good approximation since $\sysvar \lesssim \mathcal{O}(10^{-2})$, so the nonlinear interaction contributions to $\fsys$ \textit{due to each systematic acting as its own multiplicative screen} are fairly negligible, i.e.  $\fsys' \approx \left(\sum\nolimits{_{i=1}^{\Ntpl} f'_i}\right)$,
as long as the templates sufficiently capture the form of contamination: $f'_i = \alpha_i t_i$. Of course this latter condition is a basic requirement of all of the methods we describe here, one that can and should be verified through standard residual plots and other regression diagnostic techniques to ensure an appropriate contamination model for each template (though using methods that incorporate template selection criteria, such as the proposed Elastic Net, can help to satisfy this by allowing a large number of templates to be included in order to address potential higher order terms with little penalty.} Applying the multiplicative correction makes Mode Projection equivalent to the Weights model where the coefficients are derived from a simultaneous multiple regression on all the templates (such as in Refs.~\cite{Bautista_2018, ross2020completed}), but with an additional correction to debias the inferred two-point function. Thus a pixelized weights map for Mode Projection can be produced\footnote{This can be released on its own or, as with the DES-Y1 data release, as an additional column at the catalogue level. c.f. \url{https://des.ncsa.illinois.edu/releases/y1a1/key-catalogs/key-redmagic}).} as 
\be
w(\nhat) = (\shat(\nhat)+1)/(\dobs(\nhat) + 1),\label{eq:multweights}
\ee

\begin{figure}[t]
\includegraphics[width=\linewidth]{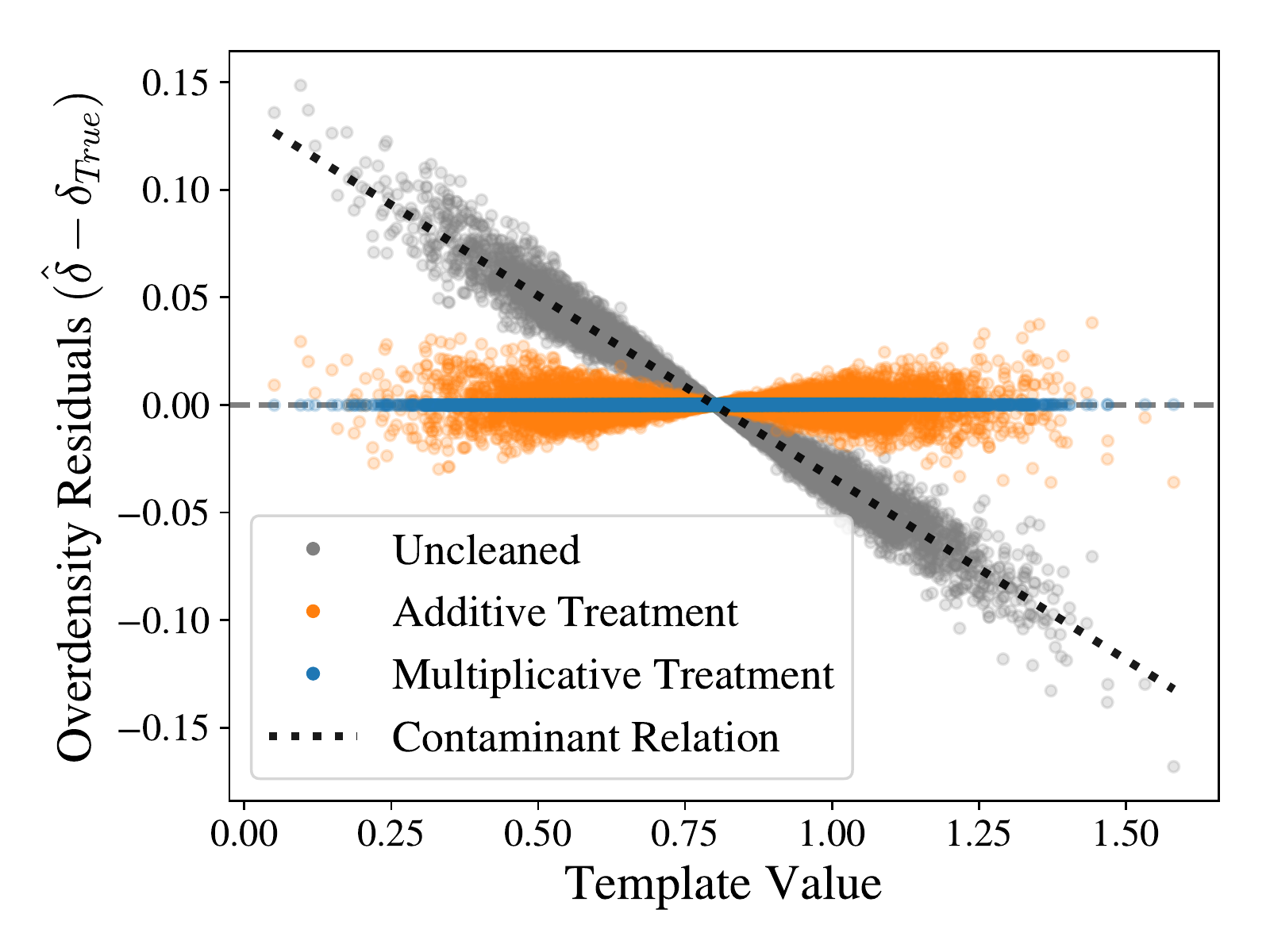}
\caption{The error in estimates of the overdensity $\delta$ in a toy Gaussian map when contaminated with a single template. Gray points indicate the pixel-based difference between the observed, uncleaned overdensity and the true overdensity when the contamination is multiplicative (additive contamination would lie directly along the dotted line). Orange points are the result when erroneously assuming the contamination is only additive. Blue points are the result when correctly treating the multiplicative component.}
\label{fig:add_mult_resids}
\end{figure}

\begin{figure}[htbp]
\includegraphics[width=\linewidth]{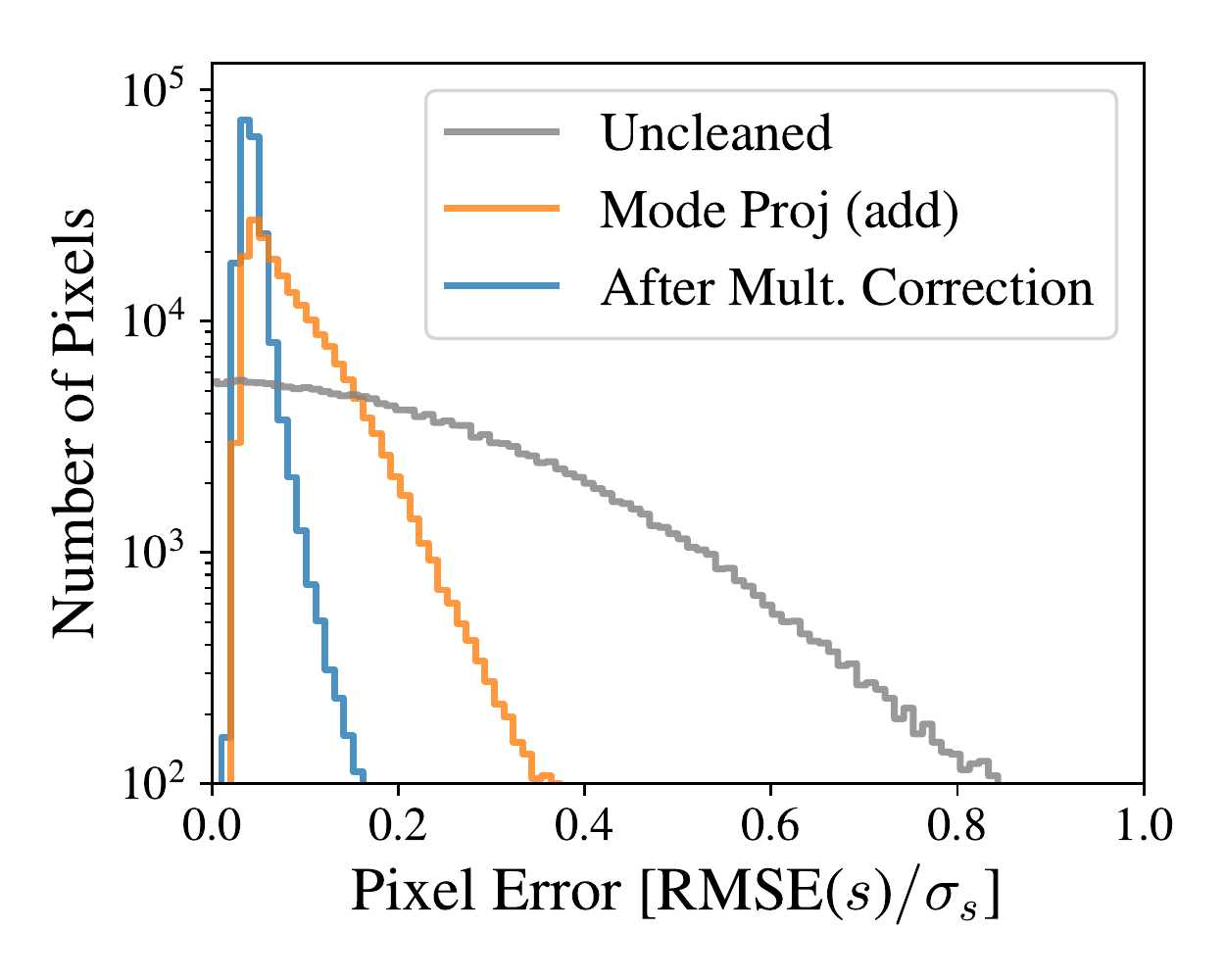}
\caption{Distribution of pixel errors before cleaning (gray), after cleaning with Mode Projection but \textit{before} multiplicative correction (orange), and \textit{after} multiplicative correction (blue). The errors have been calculated as the RMSE of each pixel across 100 cleaned mocks in our fiducial configuration of a DES-like survey as described in Sec.~\ref{sec:evalperformance}, and have been normalized to the expected dispersion from the true overdensity field.}
\label{fig:pixrmsedist}
\end{figure}

Fig.~\ref{fig:add_mult_resids} illustrates the effect of the multiplicative terms --- as well as the impact of neglecting them --- on the residuals of a map with a single linear, multiplicative contaminant.
The diagonal, dotted line shows the expected relation that would be precisely followed by a purely additive contaminant. A multiplicative  contamination adds significant scatter around this relation, shown as the gray points.
This scatter remains when the contamination is cleaned with an additive method (orange), but is effectively removed when the multiplicative component is taken into account (blue). Fig.~\ref{fig:pixrmsedist} shows how errors on the estimated overdensity field are drastically reduced when applying the multiplicative correction of Eq.~(\ref{eq:debiasedols22}) to a realistic use case with multiple contaminating systematics (see Sec.~\ref{sec:evalperformance} for details of implementation).

\subsection{Multiplicative Effect on Likelihood}
While the multiplicative term vanishes in the ensemble average, resulting in the same ensemble pixel mean as the additive-only prediction ($\langle \dobs \rangle = \fsys'$), the pixel variance is modulated:
\begin{equation}
\begin{aligned}
    {\rm Var}[{\dobs}_i] &\approx  \langle [s \gamma' (1+\fsys')]^2\rangle \\[0.1cm]
    &\approx  \langle s^2\rangle(1+\fsys')^2 \\[0.1cm]
             &=  \sigvar(1+\fsys')^2, \label{eq:vardobsmult}
\end{aligned}
\end{equation}
where for large $\Npix$ $\gtrsim 10^5$, $\langle \gamma' \rangle \approx \langle \gamma'^2 \rangle \approx 1$. The corresponding covariance between pixels is
\be
\!\!\!\!
{\rm Cov}({\dobs}, {\dobs}_j) \approx \left \langle(\gamma')^2\left[s_i(1+{\fsys}'_i)\right]\left[ s_j ( 1+{\fsys}'_j)\right]\right\rangle.\label{eq:syscov}
\ee
This is the source of the systematic-dependent scatter in Fig.~\ref{fig:add_mult_resids}, which will result in biased two-point statistics from additive methods. Because the contamination estimate is unbiased, the correction of Eq.~(\ref{eq:debiasedols22}) almost fully suppresses this variance, but the multiplicative terms also impact the likelihood when performing the regression. In pixel-space a simple fix would be to iterate: use an initial estimate of $\langle \hatfsys \rangle$ with Eq.~(\ref{eq:vardobsmult}) to apply inverse variance weights to the maps before making a second estimate of $\langle \hatfsys \rangle$. In practice these are ``errors on the errors" and so the impacts will be subdominant to the multiplicative correction to the datavector itself.

\section{Applications}\label{sec:applications}
We can use the insights of the previous sections to propose two additional methods, as well as to estimate the errors on the cleaned map. We now describe these in turn.

\subsection{Iterative Forward Selection}
We include an iterative Forward Selection method that incorporates some of the main features of the DES-Y1 method, but adopts some of the simplifying assumptions of Mode Projection. The result is greatly simplified and easier to implement than the full DES-Y1 method. 

We keep the core of the template selection algorithm, but modify the fit procedure and significance criterion to eliminate the need to generate mocks. We do this by adopting the same implicit assumptions of Mode Projection: that pixels are uncorrelated and have equal variance. This allows for an analytical solution for the best-fit parameters $\theta=\{m_i, b_i\}$ and their covariance ${\rm Cov}_{\theta}$ for each template, which we obtain using \texttt{numpy.polyfit}\footnote{To estimate ${\rm Cov}_{\theta}$, \texttt{numpy.polyfit} assumes a diagonal Gaussian covariance of the pixels, scaled so that the best-fit model has reduced $\chi^2$ of $\chi^2_{\rm red} = \chi^2/(\Npix - 2) = 1$}. 
We then adopt a simplified significance criterion of $\Delta\chi^2_{\rm FS} = \theta^T [{\rm Cov}_\theta]^{-1} \theta$, and use the same stopping threshold as the DES-Y1 method.\footnote{As with the DES-Y1 method, this method can suffer from a lack of convergence when the theshold is low, where chance correlations between the signal realization and templates result in a loop of the same series of templates being repeatedly reweighted. We  adopted a limit of $10\times\Ntpl$ reweightings for each signal realization before breaking the loop and using the resulting signal estimate as is. This occurred occasionally and at very low thresholds, with no discernible effect on the estimated maps or power spectra.}

This Iterative Forward Selection method is a fast and simple method that incorporates some of the key aspects of the DES-Y1 method, the iterative weighting and template selection, while avoiding the most computationally expensive parts, the generation of mocks. We expect some loss of precision by not including a covariance matrix in the fitting step, but on the other hand to gain some precision by not having to bin pixels, so this method can help to benchmark the importance of including the covariance matrix in a DES-Y1-like method.

\subsection{Elastic Net}\label{sec:EN}
We also propose a method that closely mimics Mode Projection but incorporates template selection, thereby reducing the impact of overfitting when the template library is large.\footnote{See \citet{Leistedt:2014wia} for an alternative approach that pre-selects templates for projection using a $\chi^2$ threshold.} Having shown that Mode Projection is equivalent to linear regression, we adopt a regression method specifically designed to automatically select predictors based on the data. 

This selection is accomplished by modifying the  \textit{Loss} function that is optimized when fitting, which is equivalent to applying a prior to the template coefficients and finding their maximum a-posteriori (MAP) estimate. Specifically, instead of finding $\hat\alpha$ that minimizes the square of the residuals ($||\dobs - T\alpha||^2$), we instead minimize
\be
{\rm Loss} = \frac{1}{2\Npix}||\dobs - T\alpha||_2^2 + \lambda_1 ||\alpha||_1 + \frac{\lambda_2}{2}||\alpha||_2^2, \label{eq:loss}
\ee
where 
\be
||\alpha||_1 = \sum_i^{\Ntpl}|\alpha_i|
\ee
is the L1-norm of $\alpha$, and
\be
||\alpha||_2 = \left(\sum_i^{\Ntpl}|\alpha_i^\dag \alpha_i|\right)^{\frac{1}{2}}
\ee
is the usual vector L2-norm of ${\alpha}$. Here $\lambda_1$ and $\lambda_2$ are hyperparameters that are tuned from the data, which we now discuss in turn:

\begin{enumerate}
    \item The L1-norm term incentivizes sparsity in $\alpha$ by penalizing non-zero coefficients of templates, thus naturally performing template \textit{selection}. This is useful because the number of templates in modern surveys can be enormous --- e.g. \citet{Leistedt:2014wia} produce $\sim3700$ templates for their analaysis of SDSS quasars --- and so it is common to pre-select only a handful to use, for fear of removing true signal. Since we don't know \textit{a priori} which templates  are contaminating, the incorporation of an automated selection scheme enables a more agnostic, data-centric approach to cleaning a large library of templates, while mitigating the risk of overfitting.
    The use of this penalty term in isolation (i.e.\ setting $\lambda_2=0$) is often called the Least Absolute Shrinkage and Selection Operator (LASSO)\cite{foster1994}, and has a Bayesian interpretation of applying a zero-centered Laplace prior on the elements of $\bm{\alpha}$, with a width $\propto 1/\lambda_1$ (see e.g. \cite{Starck_2013_bayes} for a discussion). L1 priors to induce sparsity have been used in a variety of astrophysical problems, such as for source separation in cosmic microwave background analyses \citep{bobin_sourcesep2018, Bobin_2013, wagnercarena2019novel_sourcesep} or in reconstructing mass maps from weak lensing data \citep{Leonard_2014, Lanusse_l1_2016, jeffrey_des_glimpse_2018MNRAS.479.2871J}. 
    \item The L2-norm term helps address collinearity (i.e.\ correlation) between template maps which, when present, can cause the matrix $T^\dag T$ to be ill-conditioned and the variance of contamination estimates to be large. When it is the only additional penalty term (i.e.\ $\lambda_1~=~0$), this is often called Ridge Regression, or Tikhonov Regularization. It is straightforward to show that, from a Baysian perspective, this method is equivalent to placing a zero-centered Gaussian prior on the elements of ${\alpha}$, with a width $\propto 1/\lambda_2$.
\end{enumerate}

Since each penalty term addresses a different issue with standard regression, it is not uncommon to combine them, as proposed by \citet{zouENet2005}, in a method known as the ``Elastic Net". We use the \texttt{scikit-learn}\cite{scikit-learn} implementation, \texttt{ElasticNetCV}, with a hyperparameter space of $\lambda_1/(\lambda_1+\lambda_2) \in \{0.1, 0.5, 0.9\}$ and 100 values of $(\lambda_1+\lambda_2)$ spanning three orders of magnitude, which are automatically determined from the input data (the default setting). 
We use 5-fold cross-validation to determine the best $\lambda_1$ and $\lambda_2$, trained on a random selection of $30\%$ of the input map pixels. 

In this 5-fold cross-validation scheme, the training sample ($30\%$ of the map) is itself partitioned into five equal subsamples. 
For each combination of hyperparameters, one subsample is withheld for validation, while the other four are used to train the model by minimizing Eq.~(\ref{eq:loss}). The mean squared error (MSE) of the validation sample is then computed and stored (i.e.\ the first term in Eq.~(\ref{eq:loss})). One of the four training subsamples is then withheld as the new validation set, and the process is repeated until each of the five subsamples has been used exactly once for validation, with their average MSE used to characterize the goodness-of-fit for the given set of hyperparameters $\lambda_1$ and $\lambda_2$. 

\begin{figure}[htbp]
\includegraphics[width=\linewidth]{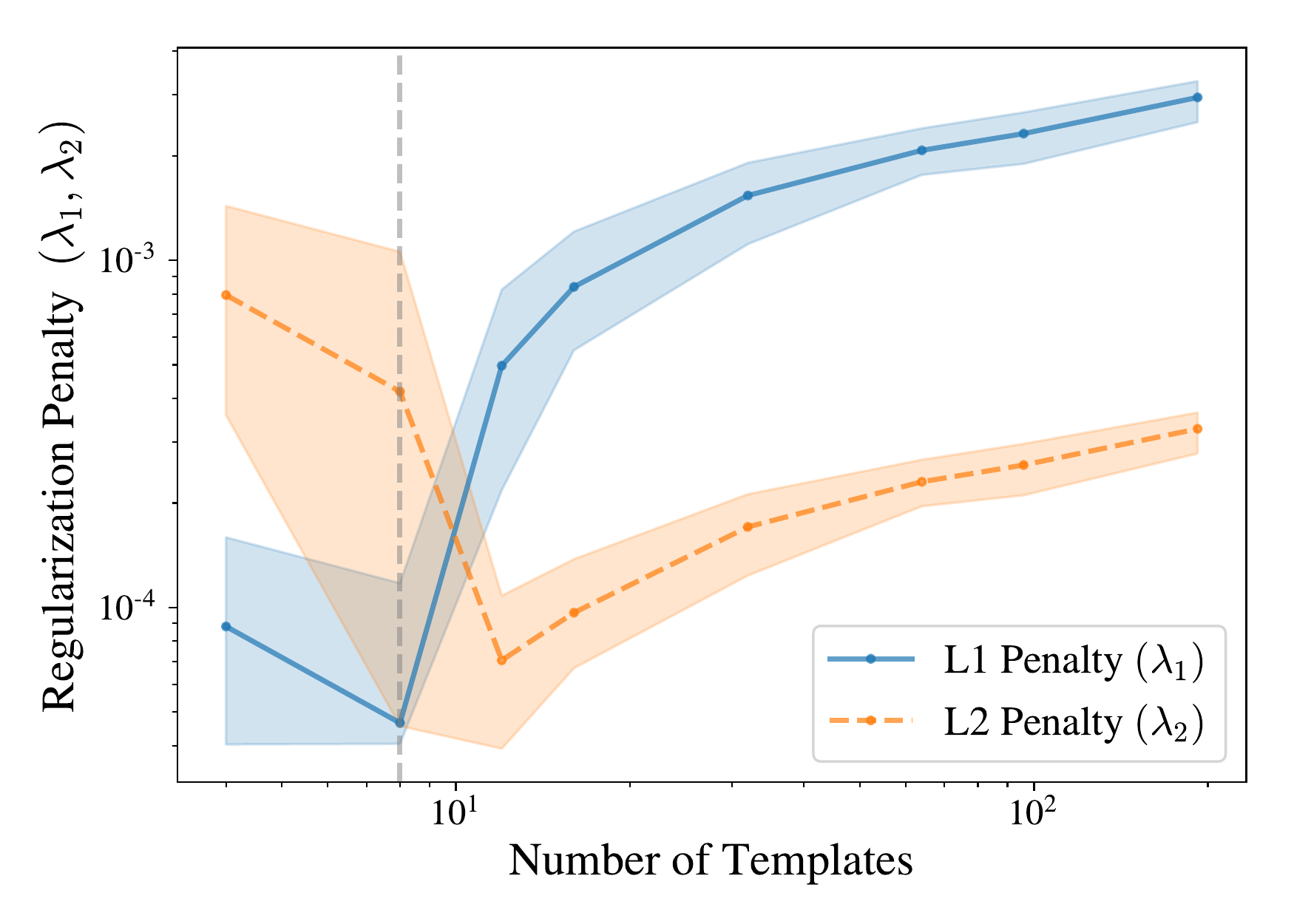}
\caption{Best-fit L1 and L2 penalty coefficients in the regularization technique described in Sec.~\ref{sec:EN}, as a function of the number of templates used for cleaning, $\Ntpl$ (new signal and template maps are generated at each value of $\Ntpl$). In all cases, 12 templates are contaminating the observed data (vertical dashed line). The importance of the L1 penalty, facilitating template selection, becomes increasingly important as more templates are included for cleaning. Lines and shaded region indicate the median and central $68\%$ probability mass of 50 mocks at each $\Ntpl$ for the central bin of our fiducial DES-like survey. Here, $\rhotpl=0.2$ within template groups, though plots for other $\rhotpl$ look similar. See Sec.~\ref{sec:evalperformance} for details of implementation.}
\label{fig:l1l2-xcorr}
\end{figure}

Setting $\lambda_1 = \lambda_2 = 0$ reduces to OLS regression and hence to the pseudo-$\Cl$ Mode Projection method, while  sampling extreme values for the relative weight of the L1 vs. L2 penalty allows for the effective use of only one of the penalty terms, if preferred by the data. The use of cross-validation on a subset of the map allows the data to dictate which model is most appropriate, with minimal risk of overfitting. We illustrate the utility of this in Fig.~\ref{fig:l1l2-xcorr}, which shows how the cross-validation scheme naturally increases the L1 penalty when fitting for more (\textit{un}contaminating) templates. We found that the L2 penalty became increasingly important when the correlation between templates increased beyond $\rhotpl\gtrsim0.9$.

\subsection{Map errors} 
\label{sec:pixelerrors}
We can use the regression framework to gain insight into how errors in the estimated overdensity map are distributed across pixels. This aids the propagation of map errors in cross-correlation studies and summary statistics beyond the two-point functions, as well as helps to identify regions that may benefit from masking.

For simplicity, we assume additive contamination and correction and ignore higher-order terms:
\be
\dadd = s + \fsys = s + T\alpha
\ee
The estimated contamination amplitude is then 
\begin{align}
\hatalphamp &= (T^\dag T)^{-1} T^\dag \dadd\\
            &= \alpha + (T^\dag T)^{-1} T^\dag s
\end{align}
such that our signal estimate is 
\begin{align}
\shat_{mp} &= \dadd - T\hatalphamp \\
            &= s - T(T^\dag T)^{-1} T^\dag s\\
            &\equiv (I - H) s
\end{align}
where the matrix $H\equiv T(T^\dag T)^{-1} T^\dag$ is often called the `Hat' or `Projection' matrix in the statistics literature. 
Then 
\be
{\rm Var}[(\shat_{\rm mp} - s)_i] = {\rm Var}[(Hs)_i] = [H{\rm Var}[s]H^\dag]_{ii}
\ee
If we make the assumption that the signal covariance is diagonal, then ${\rm Var}[s] \approx \sigvar I$ and 
\be
{\rm Var}[(\shat_{\rm mp} - s)_i] \approx \sigvar (HH^\dag)_{ii} = \sigvar H_{ii} \label{eq:addrmsepred}
\ee
where we have used the fact that H is both Hermitian and idempotent so that $HH^\dag = HH = H$.

Despite a number of simplifying assumptions and the fact that some of the methods only fit for some of the templates, we find that with the exception of Template Subtraction, $H_{ii}$ is a remarkably good predictor\footnote{Note that $H_{ii}$ only requires the diagonal elements of $H$, which are far more tractable to calculate than the full $\Npix \times \Npix$ matrix.} of how the errors in the overdensity estimates are distributed for all the methods. The errors arise from removing real signal during the cleaning process, with $H_{ii}$ as a measure of how susceptible pixel $i$ is to such overcorrection. This also indicates that while to first order all correlation with templates is removed from the estimated overdensity field, the templates remain imprinted on the map through their absence; there is missing signal in precisely their spatial configuration.  

Intuitively, $H_{ii}$ as a distance measure of pixel $i$ from the center of mass of other pixels in the $\Ntpl$-dimensional space spanned by the templates. This is sometimes referred to as `leverage', as pixels with higher $H_{ii}$ have larger impact when performing a regression.\footnote{This phenomenon is very familiar from the simple case of fitting a 1D line to a scatter of 2D points $\{x, y\}$, where the best-fit line is `pulled' preferentially to points that lie farther from $\bar{x}$.} This can be seen by observing that the estimated systematic field can be written as
\be
\hatfsys = H \dobs
\ee
such that the leverage
\be
H_{ii} = \frac{\partial \hatfsys^{(i)}}{\partial \dobs^{(i)}}
\ee
encodes the sensitivity of the contamination estimate to an observed over- or underdensity at pixel $i$. Because pixels with high leverage can have an outsized effect on the estimated contamination, we expect leverage to be a useful tool for identifying potentially problematic pixels that should be masked before cleaning, in addition to providing error estimates for those pixels that remain.

It is straightforward to derive the mean leverage value as 
\be
\bar{H}_{ii} \leq \Ntpl/\Npix,
\ee
with the equality holding if $T$ is full rank, since 
\be
\sum_i^{\Npix}~H_{ii}~=~{\rm Tr}(H)\leq \Ntpl,
\ee
providing a basis on which to determine extreme leverage values.

\begin{figure*}[htbp]
\includegraphics[width=\linewidth]{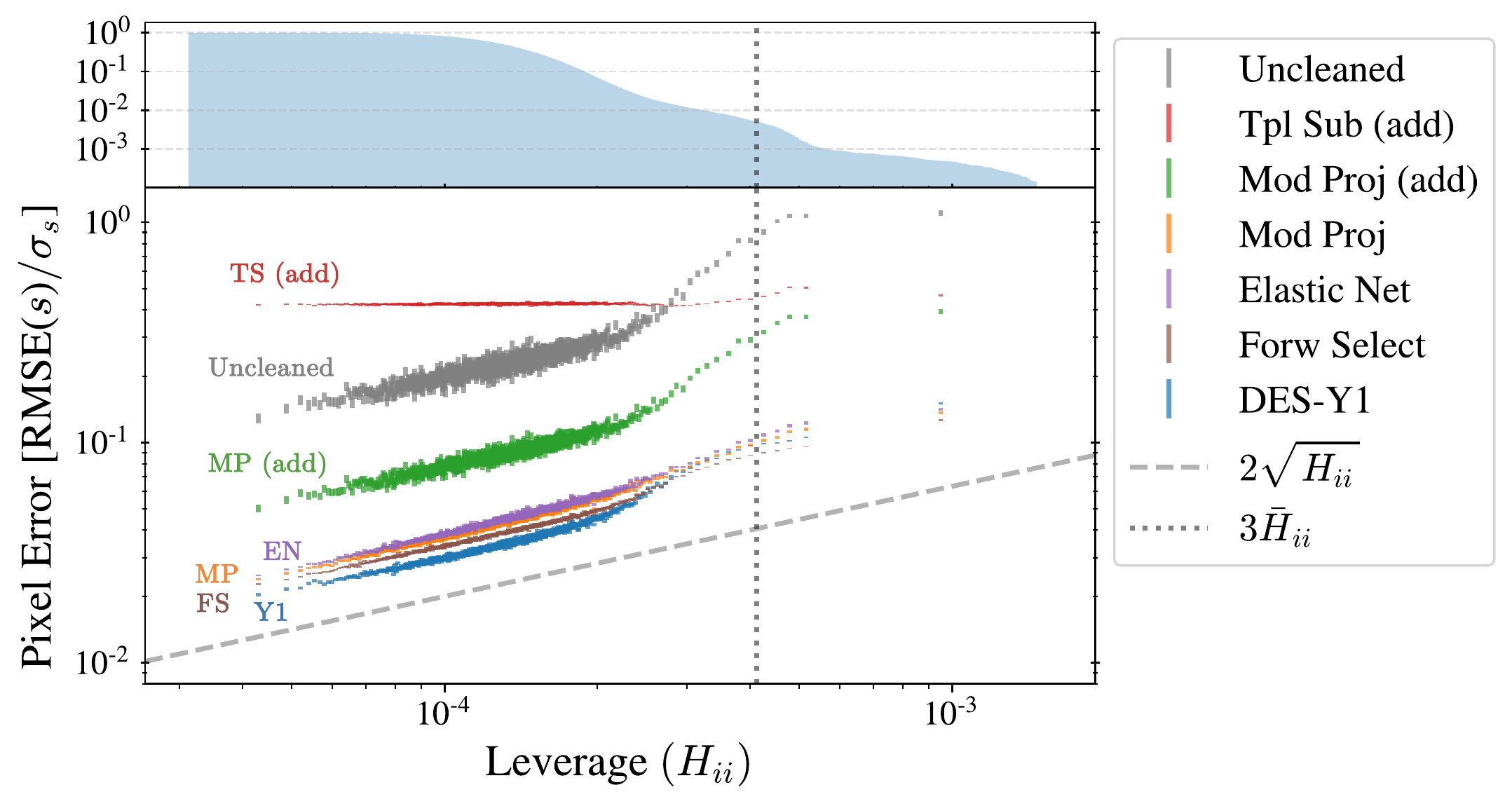}
\caption{Root-mean-square error of  pixel overdensity estimates, normalized to expected dispersion from the true overdensity due to cosmic variance, vs.\ pixel leverage for  100 signal realizations. The vertical axis shows the standard error across pixels in 1000 equal-sized bins (in this case $~197$ pixels per bin at \texttt{Nside}=128). 
The error in both observed and estimated overdensity scales as roughly $\propto H_{ii}^{1/2}$ for all methods (dashed line, to guide the eye). The dotted vertical line indicates a commonly used threshold of 3$\times$ the mean leverage across pixels to identify pixels that may have an undue impact on regression fit parameters. The histogram in the top panel indicates the number of pixels at a given leverage. $\Ntpl=27$, $\Nsys=11$, $\sysvar=0.01$.}
\label{fig:pixerrhii}
\end{figure*}

The main panel of Fig.~\ref{fig:pixerrhii} shows the RMS error (RMSE = $\sqrt{\langle(\shat - s)^2\rangle}$) of each pixel computed over 100 cleaned DES-like mock maps plotted against leverage $H_{ii}$ from 27 cleaning templates, 11 of which are contaminating. Pixels are grouped into 1000 bins of ~197 pixels, according to their leverage value, and we show the mean and standard error of the RMSE for each bin. We see that pixels with low leverage value have much smaller error in the estimated overdensity map, and that the error goes roughly as $\propto H_{ii}^{1/2}$ (diagonal dashed line), as predicted by Eq.~(\ref{eq:addrmsepred}). The Template Subtraction method is an exception to this trend likely because the regression happens in a different space, at each harmonic separately, and so does not relate cleanly to the pixel leverage.\footnote{In principle, one could construct the analogous leverage quantity $H_{\ell m}  = t_{\ell m}[\ClTT]^{-1}t^\dag_{\ell m}$ in harmonic space for the analysis of errors in $\hat{s}_{\ell m}$, which may be useful for cross correlation analyses in harmonic space.}

The top panel of Fig.~\ref{fig:pixerrhii} shows the fraction of map pixels below a given leverage (note the log scale), with the vertical dotted line indicating $3\times \bar{H}_{ii}$, which is one of two common thresholds used in statistics to flag points that may bias a regression analysis ($2\times \bar{H}_{ii}$ being the other). Here, $0.5\%$ of map pixels exceed $3\times \bar{H}_{ii}$; these pixels potentially merit further inspection or masking, as they are particularly prone to biasing the regression. The trend of the uncleaned data may be surprising, but as noted in Sec.~\ref{sec:addmultmethods}, because of the integral constraint, $\dobs$ is insensitive to a monopole in $\fsys$ and so as long as templates approximately trace the true contamination, overdensities near the mean of the templates (i.e. low $H_{ii}$) will be most accurately measured, even if contamination is greater than at other points in the map (see App.~\ref{app:monopole}).

A complementary statistic is the `Cook's distance'\cite{10.2307/1268249, doi:10.1080/01621459.1979.10481634} for each pixel, which uses $H_{ii}$ and $\shat_i$ to provide a measure of the total change in the $\shat$ map if pixel $i$ were to be masked (assuming additive contamination and correction)). Along with the leverage, we expect this to be a useful tool when performing template-based mitigation of spatial systematics and for mask creation. We leave further investigation of these as diagnostic tools, as well as generalization to the multiplicative case, to a later work. 

We next describe the fiducial survey on which we test the performance of foreground-cleaning methods.

\section{Evaluating Performance} \label{sec:evalperformance}

Our analysis is fully synthetic, with the procedure depicted in Fig.~\ref{fig:flowchart}. We compare the cleaning methods described, including results for both the standard additive Mode Projection case (denoted `Mode Projection (add.)') as well as one with the multiplicative correction from Eq.~(\ref{eq:debiasedols22}) (denoted simply `Mode Projection'). For the Elastic Net, we only show results that include the multiplicative correction.

We only consider full-sky maps in this paper. Extension to partial-sky surveys should be fairly straightforward, requiring the usual correction of cut-sky power spectra, but this applies equally across the full-sky spectra estimated with each method here and so we do not expect it to qualitatively change the main results. 

\subsection{Templates}\label{sec:templates}

We first describe the fiducial set of templates that we use, for both contamination and cleaning purposes. We adopt several classes of templates in order to span a range of possible contaminants and their spectral behavior. In most cases, we use multiple templates of the same class by generating Gaussian realizations of maps from the same theoretical power spectrum. The classes of template we use are

\begin{itemize}
    \item $\Cl \propto (\ell+1)^{0}$ (white noise)
    \item $\Cl \propto (\ell+1)^{-1}$
    \item $\Cl \propto (\ell+1)^{-2}$
    \item $\Cl \propto \exp{[-(\ell/10)^2]}$
    \item `a `Cat-scratch" map, with 128 horizontal stripes to model a basic scanning pattern and/or differences in depth due to overlapping tiles
    \item a 2D Gaussian ``spot" map
    \item a E(B-V) extinction map, with dependence on latitude removed.
\end{itemize}
The last three correspond to static maps which do not change throughout the analysis. We use the full-sky E(B-V) map\footnote{\url{https://wiki.cosmos.esa.int/planckpla/index.php/CMB_and_astrophysical_component_maps#The_.5Bmath.5DE.28B-V.29.5B.2Fmath.5D_map_for_extra-galactic_studies}} from Planck \cite{Abergel:2013fza}, but since this is dominated by emission near the galactic plane, which LSS surveys typically avoid, we reweight the map to remove its major latitudinal dependence. 

We normalize the individual templates to the same overall variance, and construct a total systematic map as a product of some or all of the individual template maps:
\begin{align}
1 + f_{\rm sys} &= \prod_{i=1}^{\Nsys}(1+\alpha_i t_i)
\end{align}
Note that this model can generally encompass contamination to any polynomial order simply by including templates that are products of others (e.g. $t_{\rm new} \equiv t_i^2$), and incrementing $\Nsys$ accordingly. Similarly, nonlinear contamination can often be made linear through an appropriate transformation of the template map.\footnote{E.g. \citet{Elvin-Poole:2017xsf} fit linear models to the square root of exposure time and sky brightness, based on how how they contribute to the depth map.} 
This total systematic map is then scaled to a desired overall map variance $\varsys$, thus determining the overall contamination field $\fsys$. We use a fiducial level of contamination of $\varsys=0.01$, as we found this to produce fluctuations similar to those seen in the DES-Y1 data \cite{Elvin-Poole:2017xsf}; this corresponds to an RMS error on $\delta$ of $\sim10\%$. Changing the level of contamination $\varsys$ did not significantly alter our results. 

We perform the contamination and cleaning procedure shown in Fig.~\ref{fig:flowchart} on each redshift bin and for each cleaning method over many sky realizations, and plot the mean and central $68\%$ probability mass of the relevant quality statistic. We use the same set of templates and total systematic map for across redshift bins and sky realizations, but generate a new set for each unique combination of parameter choices (e.g. level of cross-correlation between templates, number of templates used, etc.) in order to minimize any effects from specific template realizations. 

We use \texttt{CLASS} \cite{Lesgourgues:2011re} to compute theoretical galaxy clustering power spectra for a mock LSS survey, including contributions from redshift-space and Doppler distortions and lensing. We found gravitational potential terms to contribute $\lesssim 1\%$ to the resultant $\Cl$ for $\ell > 7$ but increased computation time by an order of magnitude, so we neglect them. Since we find the cleaning procedures are not strongly sensitive to the signal power spectrum, this should not impact our results. We then use \texttt{Healpy}\cite{healpy} to generate full-sky Gaussian realizations of large-scale structure overdensity ($\delta \equiv \delta\rho/\rho$) maps for each redshift bin with $\texttt{NSIDE}=128$. We compare the impact of using lognormal maps in App.~\ref{app:lognorm}, finding it does not change our results.

\subsection{Cosmological model and simulated survey} \label{sec:modsurveys} 

We assume a standard $\Lambda$CDM cosmological model with one species of massive neutrino and parameter values from best-fit Planck 2018: 
$\{\Omega_c,\Omega_b,h,n_s,\sigma_8,m_\nu/{\rm eV}\}=\{0.26499,0.04938,0.6732,0.96605, 0.8120, 0.06\}$. 
Given the precise parameter constraints from current probes, the dependence of our results on cosmological parameters is expected to be very minimal.  In contrast, the choice of the parameter set to be \textit{determined} from the survey may be highly dependent on the residual systematics. 

In general for comparing the methods, the exact form of the galaxy power spectra is not very consequential, so we use a fiducial survey comparable to the completed Y5 Dark Energy Survey, for which a realistic level of contamination can be estimated based on existing data. We assume the number density distribution of galaxies to be in the form
\be\label{eq:desdndz}
\frac{dn}{dz} \propto \left(\frac{z}{z_0}\right)^{\alpha}\exp{\left[-(z/z_0)^{\beta}\right]},
\ee
where $z_0=0.55$, $\alpha=2.65$, and $\beta = 3.34$. 
We assume five redshift bins centered at redshifts $\{0.225, 0.375, 0.525, 0.675, 0.825\}$, with galaxy bias of $\{1.4, 1.6, 1.6, 1.95, 2\}$, respectively, and containing galaxies with Gaussian redshift dispersion of $\sigma_{\rm z}=0.05$. These values were chosen to closely approximate the \texttt{REDMAGIC} redshift distribution given in \citet{Elvin-Poole:2017xsf}. 

We choose to work primarily in harmonic space. Therefore, starting with some map with overdensity $\delta\equiv \delta N/N$, where $N$ is the galaxy count over some patch, the expansion in spherical harmonics gives
\begin{equation}
    \delta(\nhat)=\sum_{\ell=0}^\infty\sum_{m=-\ell}^\ell\alm \Ylm(\nhat),
\end{equation}
and the angular power spectrum is given by
\begin{equation}
    \Cl = \sum_{m=-\ell}^\ell\frac{|\alm|^2}{2\ell+1}.
\end{equation}
Because we are working in the full-sky limit, all well-known estimators of power return the same result, so here we make use of the \texttt{anafast} and \texttt{alm2cl} functions in \texttt{Healpy}. To more accurately account for the cosmological impact of the cleaning methods on data from a DES Y5-like survey, we divide the assumed sample variance $\sigma^2_{\Cl}$ by a factor of $\fsky=0.116$. 

We add shot noise to the theoretical power spectrum as $\Cl \rightarrow \Cl + \bar{n}^{-1}$, with $\bar{n}=1.5\times 10^8$, but this is negligible at the large scales we work with ($\ell \le 350$).
We are primarily interested in studying the systematic impacts of cleaning (or not) using spatial templates, so it is reasonable to focus on cases where the signal-to-noise is large (i.e.\ shot noise is negligible).\footnote{Shot noise may have the effect of (1) rendering the 
the regression residuals more diagonal in pixel-space (or flattening them in harmonic space), which could actually improve the regression procedure, and/or (2) introduce significant skewness in the distribution. We would expect the impacts of these to be similar to those of prewhitening the data or using lognormal mocks, and so based on our results in Apps.~\ref{app:lognorm} and \ref{app:prewhitening}, we do not expect shot noise to significantly impact on our findings.}

 
\section{Simulation Results}\label{sec:results}
To compare methods, we compare the fidelity of the cleaned data products to the truth, either at the map level or at the level of the power spectrum, rather than look for cosmological-parameter biases.  We do this for a few reasons: (1) the map and power spectrum are more general, being independent of (but easily mapped to) any specific cosmological model one wants to test, or summary statistic one wants to use; 
(2) while we primarily study applications to galaxy clustering data here, the methods themselves are quite general and can easily be applied to other data sets for which one has tracers of potential contamination, such as shear or convergence maps; (3) galaxy clustering \textit{alone} leads to relatively weak cosmological constraints and is rarely used on its own to constrain cosmology.

We therefore limit ourselves to investigating biases in data space and leave the investigation of impacts on cosmological constraints to a later work when weak lensing data can be incorporated in a more realistic fashion. At this stage, the test bed is sufficiently representative to compare foreground-cleaning methods in a manner to inform future LSS analyses.

\subsection{Characterizing Performance}\label{sec:mapstats}
We first study the impacts of the different methods on the estimated maps and power spectra for a single configuration and compare the residual biases of each. For this fiducial comparison, we generate 50 mocks for each redshift bin and contaminate them with 11 systematics, two from each of the four Gaussian classes, plus the three static templates. We construct a template library that contains the contaminating templates, plus four additional realizations from each Gaussian class, for a total of 27 cleaning templates. Each method uses this library to produce estimates of the overdensity field and power spectra. 

\begin{figure}[htbp]
\includegraphics[width=\linewidth]{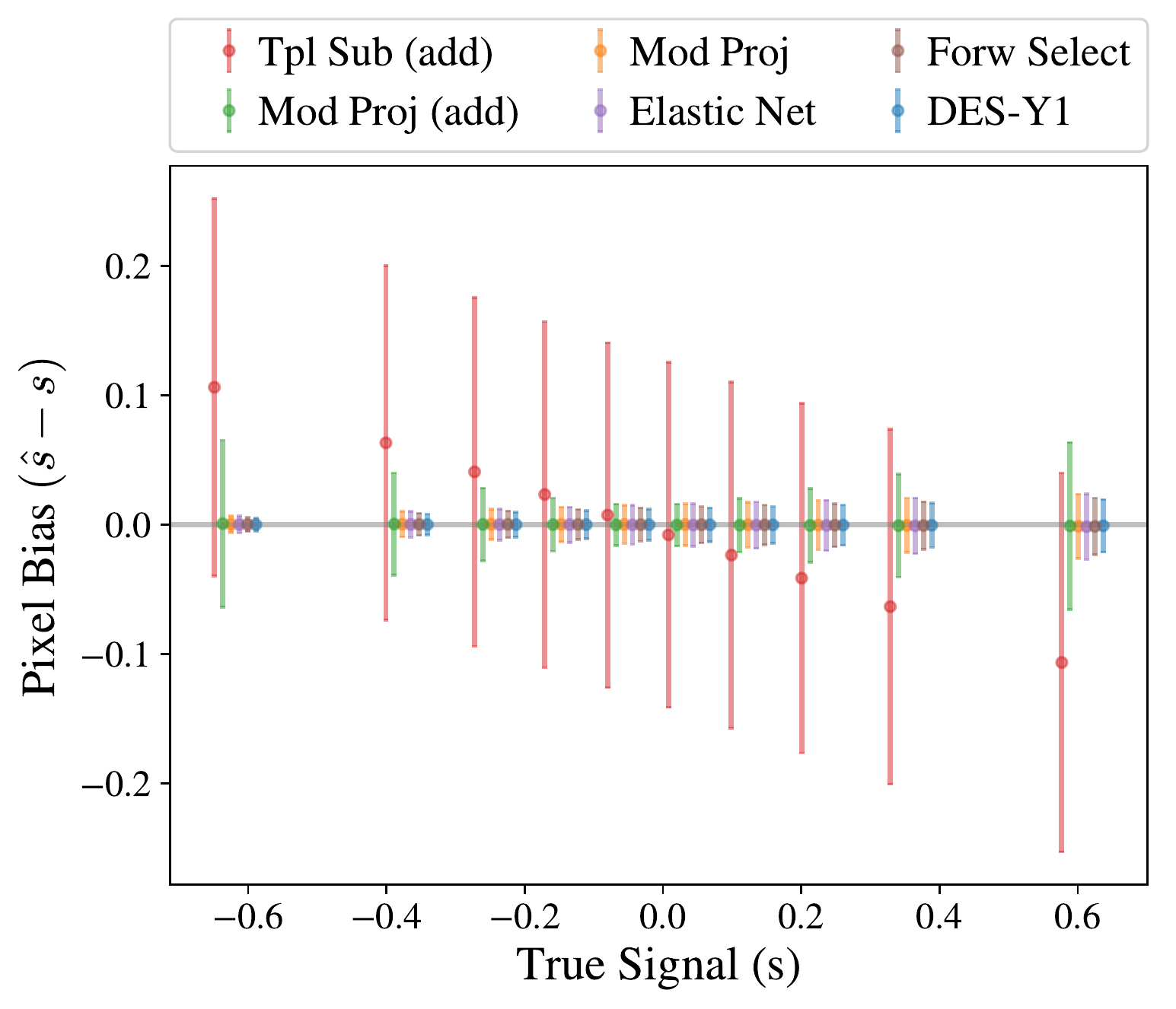}
\caption{Error in overdensity estimates for different cleaning methods, binned in deciles of the true overdensity and with points offset for clarity. Error bars show the standard deviation in each bin.  Overcorrection at the map-level is only significant for Template Subtraction, which under-estimates the magnitude of both peaks and voids, while other methods are very close to unbiased. See text for details.}
\label{fig:pixelbias}
\end{figure}

We show map residuals of each cleaning method for the lowest redshift bin in Fig.~\ref{fig:pixelbias}, where the residuals are binned into deciles of the true overdensity. Results for other redshift bins are similar. From left to right, in approximate order of performance, the figure shows the Template Subtraction method (red), Mode Projection without (green) and with (orange) multiplicative correction, the Elastic Net method (purple), Forward Selection (brown) and the DES-Y1 method (blue).

The overcorrection of Template Subtraction is evident, with density fluctuations consistently \textit{under}-estimated (i.e.\ peaks and voids are both less extreme than they should be). The other methods are all very close to unbiased with respect to the true overdensity field, with bias of the mean $\lesssim 0.001$ for each bin. The multiplicative methods show significantly reduced within-bin scatter (i.e. smaller error bars) compared to the additive ones --- the additive Template Subtraction and Mode Projection methods (leftmost, red and green) have typical errors in the overdensity of $\sigma_s \sim 0.1$ and $\sigma_s \sim 0.01 - 0.05$, respectively, compared to the errors of $\sigma_s \sim 0.005 - 0.02$ for the multiplicative methods. This suggests that applying the multiplicative correction results in significantly improved map estimates, making them excellent candidates for map-based analyses, such as as counts-in-cells or density-split statistics.

While the signal estimates are unbiased (with the exception of Template subtraction), the errors of the additive methods increase near extremes of the density field. This is similar to the result in Fig.~\ref{fig:pixrmsedist}, which showed larger errors at extreme template values, in part for the same reasons. Both Figs.~\ref{fig:pixerrhii} and \ref{fig:pixelbias} indicate a clear stratification of the methods, with the methods that fail to treat the multiplicative component of contamination showing significantly larger error.

\begin{figure*}[t]
\includegraphics[width=\linewidth]{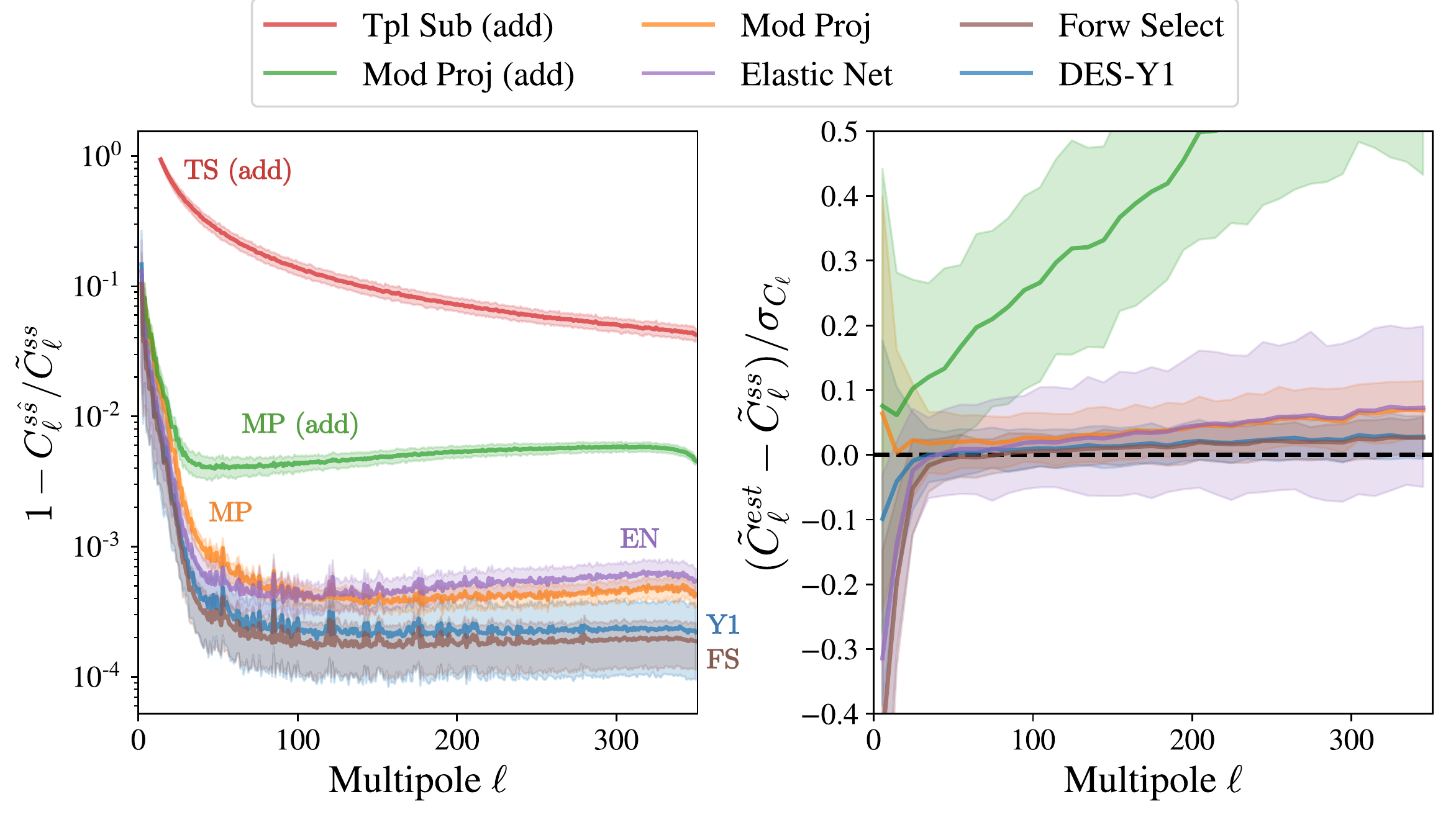}
\caption{\textit{Left}: Error in map reconstruction for each method as a function of multipole $\ell$ in a DES-like survey, shown as the deficit in correlation at each multipole between the true and cleaned maps $(1-C_{\ell}^{s \shat}/\Clss)$. A perfect reconstruction corresponds to 0, whereas pure noise corresponds to 1. For all methods (except perhaps Template Subtraction), the cleaned map is a good approximation of the true map for cross-correlation purposes, especially at scales $\ell \gtrsim 30$. \textit{Right}: Error in power spectrum estimation, shown as the residual angular power relative to sample variance $(\Clest - \Cltrue)/\sigma_{\Cl}$ in bins of $\Delta\ell=10$. Solid lines indicate means of cleaning performed on 50 signal realizations of each bin and shaded regions indicate the central $68\%$ probability mass of the 250 total realizations. The multiplicative correction applied to Mode Projection removes most of the bias of the method (green to orange). 
Here we use 27 templates, of which 11 are contaminating the data.}
\label{fig:mapclean}
\end{figure*}

We also compare the maps in harmonic space. The left panel of Fig.~\ref{fig:mapclean} shows the  per-multipole performance of the cleaning algorithms as $(1-C_\ell^{\shat s}/\Clss)$ vs. the multipole $\ell$, where
\be
\Cl^{s\hat s} = \langle\slm\hat{s}_{\ell m}^*\rangle
\ee
This quantifies the fractional \textit{missing} cross-power between the the true and estimated maps, such that a perfect reconstruction corresponds to 0, and pure noise corresponds to 1 (note the log scale). This conveys the approximate level of error expected when using cleaned maps for cross-correlation studies.

All of the methods that treat the multiplicative contamination perform significantly better than the additive methods. 
The corrected Mode Projection and Elastic Net, and the DES-Y1 method all have excellent performance at $\ell\gtrsim 30$ or scales below about $0.2$ degrees on the sky, showing $\lesssim0.1\%$ error. Maps cleaned with these methods should therefore be excellent candidates for cross-correlation studies. Even the additive mode projection method performs quite well with error of $\lesssim1\%$ in this case, and as such it may be adequate for many studies.



In the right panel of Fig.~\ref{fig:mapclean} we show the error in the power spectrum estimate as the difference between the estimated (after cleaning) and true angular power in bins of $\Delta \ell=10$ and normalized to sample variance $(\Clest - \Cltrue)/\sigma_{\Cl}$, where for $\sigma_{\Cl}$ we use the standard Gaussian approximation for cosmic variance, scaled by $1/\fsky$. Unlike $C_{\ell}^{s\shat}$, this quantity is insensitive to phase-differences between the true and reconstructed maps of the map. 

To lowest order, all of the methods work well, and the residual biases are below cosmic variance for the large angles studied here (note that systematic shifts will become more significant with larger multipole bins). For mode projection, the performance is satisfactory only once it is corrected for the multiplicative bias via Eq.~(\ref{eq:debiasedols2}). We do not show Template Subtraction on the right for clarity --- its mean traces the mean for the additive Mode Projection method, but the dispersion is very large, exceeding the plot limits.

\subsection{Susceptibility to overfitting}\label{sec:overfit}

\begin{figure*}[htbp]
\includegraphics[width=\linewidth]{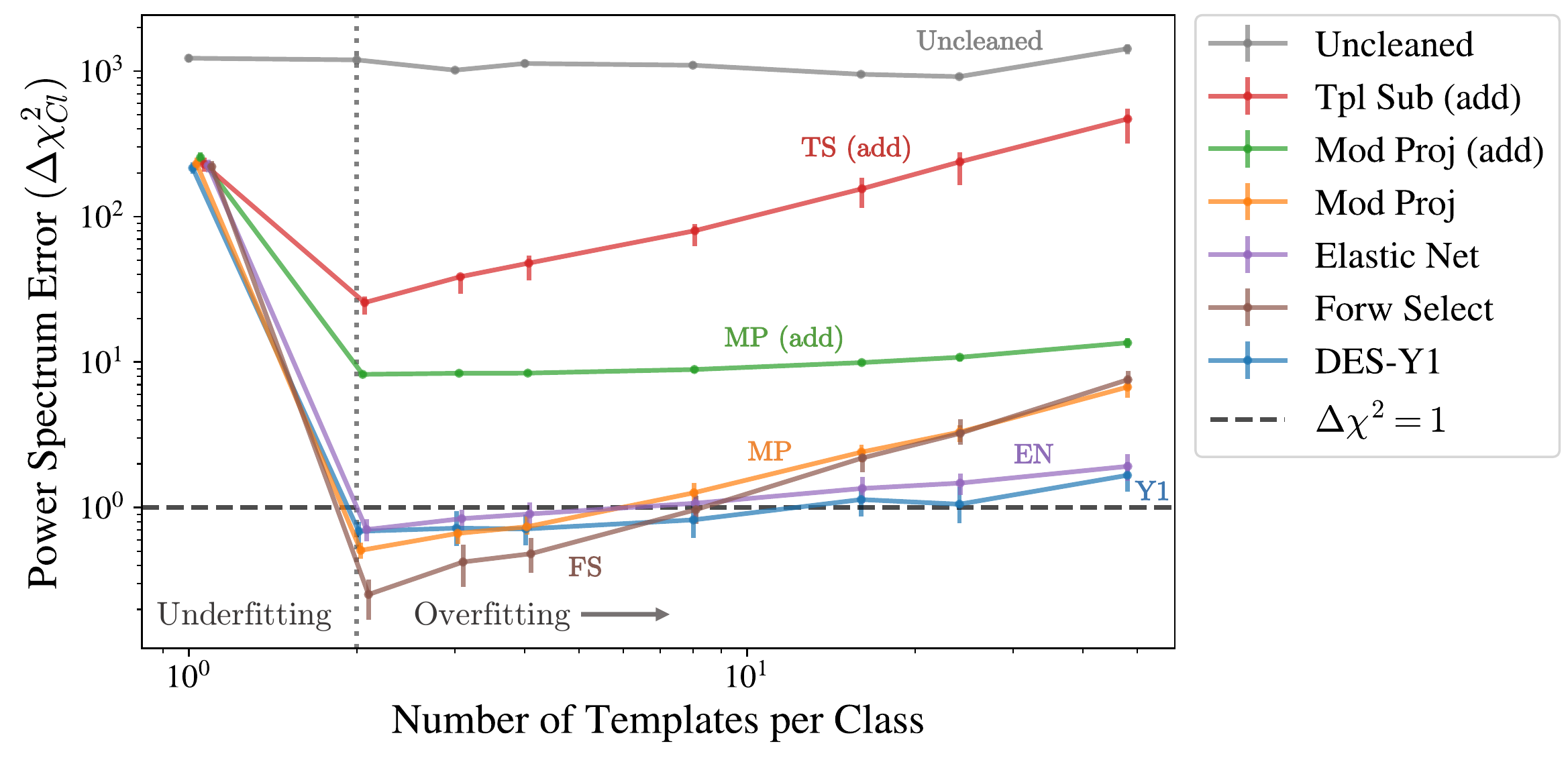}
\caption{Bias in the angular power spectrum, $\Delta \chi^2_{\rm \Cl}$, as a function of the number of templates fit to the map. We consider Gaussian templates which have a correlation of $\rho_{tpl}=0.2$ within each of the four template classes, as defined in Sec.~\ref{sec:templates}. We generate two template realizations per class with which to contaminate each signal map ($\Nsys=2$, denoted by the vertical dotted line). The templates used to perform the cleaning vary from one to 48 for each type of template spectrum, for a total of four to 196 templates. The Template Subtraction, Mode Projection, and Forward Selection methods are all mildly susceptible to overfitting --- signaled by the increase in $\Delta \chi^2_{\rm \Cl}$ for $\Ntpl>2$ --- though only Template Subtraction to a degree where it overcomes the penalty for neglecting a contaminating template ($\Ntpl=1$).  For the \textit{additive} Mode Projection method, $\dchisqcl$ is dominated by the bias from not addressing the multiplicative contribution to the power spectrum (see Fig.~\ref{fig:mapclean}, right panel), while the other methods are dominated by increased variance from chance correlations. The bias from failing to correct for the multiplicative term dominates even when fitting for $\sim200$ templates. The DES-Y1 and Elastic Net display a lesser dependence on $\Ntpl$, and so are more robust to overfitting.  See Sec.~\ref{sec:overfit} for details.}
\label{fig:nclean}
\end{figure*}

Any template-fitting model faces a challenge to neither underfit nor overfit the data. In the case of underfitting, residual contamination will be left over in the map and inferred to be signal. In the case of overfitting, a portion of the signal will be inadvertently removed from the map, having been mistaken for systematics. Additionally, increasing the number of fitted templates increases the variance of the estimated power spectrum, which will increase the error of $\Clest$ in a mean-squared sense\cite{Elsner:2016bvs}.

Mode Projection and Template Subtraction address the risk of overfitting by estimating how much signal power is lost from over-correction given the template library and scaling the power spectrum accordingly (Eqs.~\ref{eq:tsclssmult} and \ref{eq:clestmp}). In contrast, the DES-Y1 and Forward Selection methods use thresholds to limit the templates used for cleaning to only those that are most significant, an approach that was also implemented in the \textit{Extended} Mode Projection method of \cite{Leistedt:2014wia} for the QML power spectrum estimator (though as shown by \cite{Elsner:2015aga}, this comes at the cost of an unknown bias in the power spectrum). As described in Sec.~\ref{sec:EN}, the Elastic Net reduces overfitting by adding a prior on the template coefficients to reduce the number of templates used. 

While each of the methods addresses overfitting in its own way, the library of templates fed to them has in most cases already been narrowed from a much larger set of possible templates through decisions made by researchers. For example, almost all modern surveys observe any given patch of sky multiple times, resulting in multiple values for each observing condition for every pixel. To produce a scalar template map requires compressing these values into a summary statistic and, as it isn't known a priori which statistic will best capture systematic contamination of the data, multiple statistics may be computed, each corresponding to its own template (see e.g. Ref.~\cite{Leistedt:2015kka}). If just one statistic (such as the mean) is chosen as representative as is often done, there is the very real risk of discarding potential templates that more accurately capture the contamination, resulting in residual contamination, or \textit{under}fitting.

\textit{Therefore, one of the key performance metrics for these methods is their ability to handle increasing numbers of non-contaminating templates without degrading map or power spectrum estimates, and so simultaneously mitigate the risks of under- and over-fitting.}

To characterize the error in the reconstructed angular power spectrum, we use the sum of squared errors between the true and reconstructed power spectra, normalized by sample variance:
\be 
\Delta \chi^2_{\rm \Cl} = \sum_{z {\rm bins}} \sum_{\ell=\ell_{\rm min}}^{350} \frac{\left(\Clest(z) - \Cltrue(z)\right)^2}{\sigma^2_{\Clss(z)}},\label{eq:dchisqcldef}
\ee
where $\ell_{\rm min} = 2$, except for Template Subtraction where $\ell_{\rm min} = {\rm Ceil}[(\Ntpl-1)/2]$, since with $\Ntpl$ templates, all signal is removed for $\ell \leq (\Ntpl-1)/2$.

In Fig.~\ref{fig:nclean} we show $\dchisqcl$ as a function of the number of templates used to clean the maps (Fig.~\ref{fig:nclean_rmsmap} shows the same plot for map-level statistics, which demonstrate very similar behavior to $\dchisqcl$). We generate two template realizations per class with which to contaminate each signal map, and vary the number of templates used to perform the cleaning from one to 24 for \textit{each template class}. The true contaminants are always `selected first', such that $\Ntpl = \Nsys = 2$ represents correctly fitting for the two contaminating templates from each class (vertical dotted line), whereas $\Ntpl >2$ indicates the penalty for overfitting of non-contaminating templates. The error bars come from many signal realizations for the same template maps, and different template and signal map realizations are used for each value of $\Ntpl$.

Fig.~\ref{fig:nclean} demonstrates that all methods are susceptible to overfitting, as indicated by the fact that $\Delta \chi^2_{\rm \Cl}$ increases for $\Ntpl>2$, but that some are more susceptible than others. Template Subtraction and additive Mode Projection are the worst-performing methods with $\Delta \chi^2_{\rm \Cl}\gtrsim 10$ for all cases, with Template Subtraction showing a strong dependence on $\Ntpl$. Multiplicative Mode Projection and Forward Selection display approximately the same $\dchisqcl \tilde\propto \Ntpl$ scaling as Template Subtraction, whereas The Elastic Net and DES-Y1 methods show a much weaker scaling, indicating that they are much more robust to a larger number of templates.

The trend for \textit{additive} Mode Projection method indicates the importance of the multiplicative correction. Here, the error in the power spectrum does not scale with $\Ntpl$ as strongly as that of Template Subtraction or the multiplicative Mode Projection method because it is dominated by the bias from not addressing the multiplicative contribution to the power spectrum (see Fig.~\ref{fig:mapclean}, right panel), not the increased variance from a larger number of templates. The bias from failing to correct for the multiplicative term dominates the additive Mode Projection error even when overfitting by $\sim200$ templates. Were the plot to continue to the right, we would expect the error to begin to scale similarly to the other Mode Projection and Template Subtraction methods.\footnote{The multiplicative bias is not the dominant contribution for Template Subtraction because its \textit{effective} number of templates is much larger, since it performs $N_\ell$ regressions for each template.}

Another key point is that for all cases except Template Subtraction, the penalty for overfitting is dwarfed by the penalty for neglecting contaminating templates ($N=1$ on the x-axis). This suggests that the researchers should err on the side of overfitting, rather than risk removing contaminating templates from the cleaning library. This is especially true if using a method that is more robust to overfitting, such as the Elastic Net or DES-Y1 method. In sum, the DES-Y1, Forward Selection, Mode Projection with multiplicative correction, and Elastic Net methods all perform very well relative to the uncleaned case, with the Elastic Net and DES-Y1 methods being most robust to overfitting and achieving the best performance with $\Delta \chi^2_{\rm \Cl}\simeq 1$ even when $\Ntpl\gg \Nsys$.

\begin{figure*}[t]
\includegraphics[width=\linewidth]{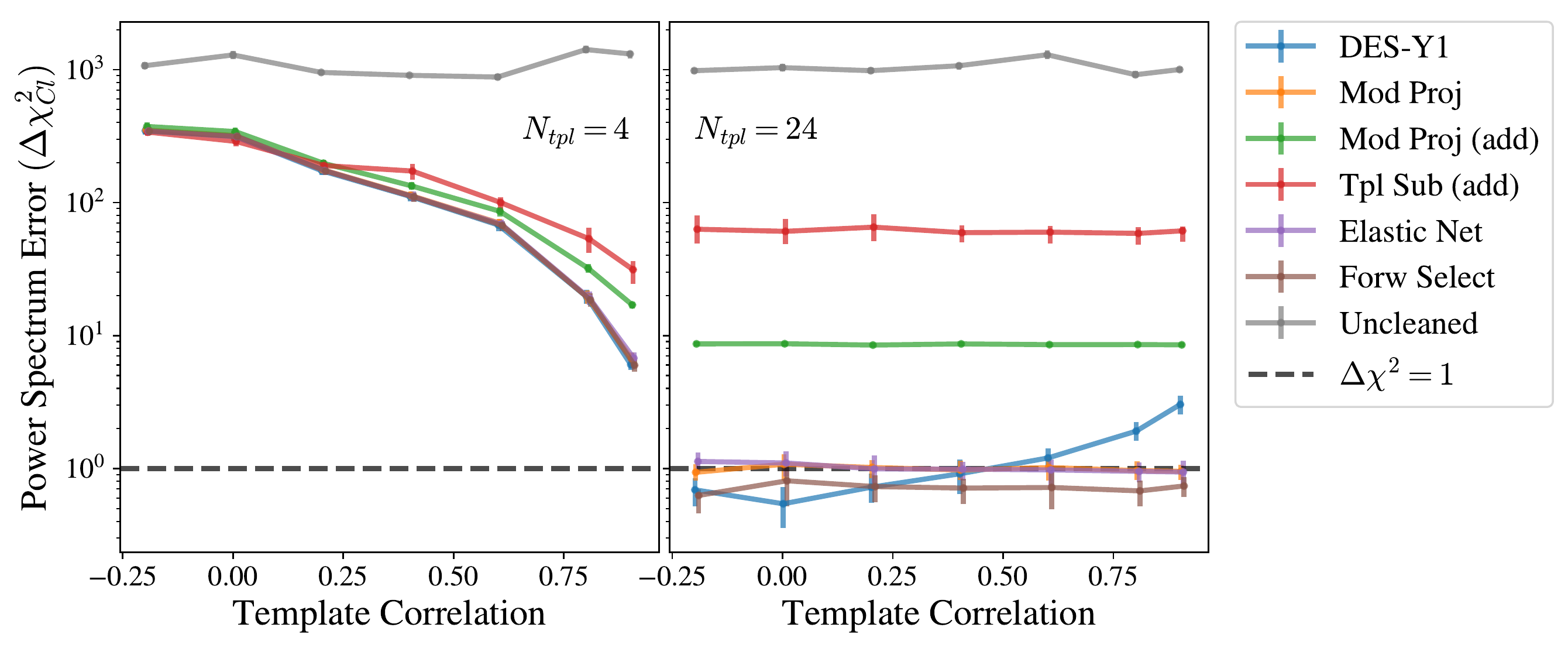}
\caption{Bias in the angular power spectrum, $\Delta \chi^2_{\rm \Cl}$, as a function of the level of cross-correlation imposed between the templates within the same class. We assumed contamination from two realizations from each of the four classes (i.e. $\Nsys=8$). The left panel assumes cleaning with only one of the contaminating templates from each class, while in the right panel we clean for four templates from each class, including the contaminating ones. Note that in the case where template correlation $\rhotpl \rightarrow 1$, the two templates are identical and it is equivalent to cleaning only for one contaminating templates, an ideal scenario. 
In the right panel we see that while the DES-Y1 outperforms others when templates are completely orthogonal, it suffers as the level of correlation between templates increases. The Elastic Net method mitigates this problem.}
\label{fig:ncleanxcorr}
\end{figure*}

\subsection{Impact of correlated templates}\label{sec:xcorr}
Real templates often have groups of templates that are highly similar to one another in their spectral behavior and/or in their correlation to one another, which we have modeled here as different template classes. 
The same tracer/property measured in different wavelength bands, or different summary statistics (e.g. the mean vs. median) for the same tracer in a multi-epoch survey are both common examples that can result in very similar templates.
We wish to investigate the impact of selecting a non-optimal template for cleaning, which only partially describes the true systematic. This could be either through the choice of a non-optimal summary statistic, or through the apriori choice of a `representative' template from a group of similar templates in order to mitigate the risk of overfitting, as is commonly done in current surveys. 

We test this by cleaning with sets of templates that have varying levels of within-class correlation. For each template class (corresponding to one of the spectra listed in Sec.~\ref{sec:templates}) we use \texttt{Healpy.synfast} to generate template realizations with off-diagonal covariance terms between templates $i$ and $j$ of
\[ 
    C_\ell^{ij} =
    \begin{cases} 
          \rhotpl \sqrt{C_\ell^{ii}C_\ell^{jj}}, &\text{if $i$ and $j$ in same class} \\
          0, &\text{if $i$ and $j$ in different classes}
       \end{cases}
\]

We only use the first four classes from Sec.~\ref{sec:templates}, which are defined by their spectrum and from which we can generate multiple Gaussian realizations with defined levels of cross-correlation.

Fig.~\ref{fig:ncleanxcorr} shows the performance of the methods when the within-class correlation between templates is varied. We again consider the case of two contaminating systematics from each of the four Gaussian template classes. The left panel shows the case where for each class we have chosen only one of the templates to clean with, deeming it ``representative" of the template group. As within-class correlation between the systematics increases, the cleaning templates are more representative and can increasingly remove more of the unaccounted-for contamination. At $\rhotpl=0.9$, the multiplicative methods are able to reduce the error to $\dchisqcl\sim6$ compared to $\dchisqcl\sim300$ for the uncorrelated case.

Despite the additional freedom of the Template Subtraction method to fit multipoles independently, it does not do a better job than the other methods of correcting for the ``unknown" systematics. The multiplicative methods have almost identical performance, with the dominant contributions to residual errors in the power spectrum resulting from the unaccounted-for systematics and, to a lesser extent, failing to treat the multiplicative term of the contamination. 

The right panel in Fig.~\ref{fig:ncleanxcorr} illustrates the other approach of including many possible templates rather than preselecting a few: we use six cleaning templates from each class: the two true systematics, plus four more that are uncontaminating, for a total of 24. We find that with the exception of the DES-Y1 method, performance of the methods is largely independent of the correlation between templates.\footnote{We found this to be true for both map-level and 2-pt reconstruction statistics, though we only show the latter here. It is not obvious from the outset that this would be the case --- Forward Selection methods are often criticized for being less reliable when predictors are correlated, though this is in the context of the more typical regression scenario where it is the predictors themselves that are of interest, as opposed to the residuals which is our focus here.
The source of the dependence of the DES-Y1 results on $\rhotpl$ is not entirely clear, but our investigations found it to be mildly impacted by both binning choices and the total monopole of systematic maps.}

Comparing the panels, even if using a high threshold of similarity of $\rhotpl = 0.9$ to discard templates, significantly more error is introduced through neglecting a contaminating template than through overfitting, so it is better to not pre-select templates solely on the basis of similarity to others and instead err on the side of too many templates rather than too few. Template Subtraction is the one exception to this, where each additional template results in $N_\ell$ additional fits. While the additional freedom does not substantially protect against unknown systematics, it does result in a much steeper penalty for overfitting from higher $\Ntpl$.

\subsection{Extensions}\label{sec:extensions}
By interpreting current LSS systematics cleaning methods in the context of regression, we have facilitated their comparison and interpretation, as well as motivated several possible extensions to them. We have explored some of these extensions in this work, such as the Elastic Net method in Sec.~\ref{sec:EN}, and the use of leverage to predict overdensity errors, but with the extensive body of regression methods, there are many more that we must leave to future work. For example, one promising avenue for regression methods that use a threshold for template selection would be to motivate that threshold by controlling the ratio of Type I (false correction) to Type II (false omission) errors in the selection process via the False Discovery Rate\cite{BH_FDR_95}, based on the relative impact of each type of error on the analysis.

We have noted individually multiple cases where the assumptions made by the methods do not hold and how they might be improved. A full treatment of these effects is beyond the scope of this paper and would include the full non-Gaussian likelihood of $P(\dobs|\hatfsys)$, including contributions from systematics, but as we show in Apps.~\ref{app:lognorm} and \ref{app:prewhitening}, the corrections from these are minor compared to the methodological differences and the improvements we suggest. Generalized linear models may be a promising compromise for future mitigation routines, preserving off-the-shelf implementation and diagnostic tools, while providing greater specificity for the likelihood and relaxing some of the tacit assumptions of Mode Projection and OLS regression.

The methods presented here are general enough to be applicable in any situation where one has an external prediction (template) for systematic contamination of observational data, and is equally applicable to spin-2 fields. The insights gained can be used to further extend linear models like the ones in this work, or inform the formulation of nonlinear contamination models, non-parametric methods, or machine learning approaches such as that of \citet{Rezaie:2019vlz}.

\section{Summary of Methods}\label{sec:summary_methods}

Here we summarize our findings about the performance of systematic-cleaning methods. 

\begin{itemize}
    \item \textbf{DES-Y1 method}: The most complicated method of the ones we studied, the DES-Y1 method resulted in some of the lowest biases in the cleaned maps. It usefully includes prior information about the covariance between pixels in the fitting procedure, albeit in a coarse way. However it is also somewhat complicated to implement, as it requires a large number of parameter choices on the part of the researcher (binning number and procedure, significance statistic and threshold, power spectrum prior) and the generation of realistic mocks. We observed some degradation of its performance as the correlation between templates increased. It is one of the two methods most robust to overfitting when using a large library of templates that are not actually contaminating the data (the other being Elastic Net).\\
    \item \textbf{Mode Projection:} The standard pseudo-$C_\ell$ Mode Projection method, as introduced in \cite{Elsner:2016bvs} and implemented in \texttt{NaMaster}\cite{Alonso:2018jzx}. We showed that it is equivalent to removing the result of an ordinary least squares regression of the observed data onto the template maps (thus providing a map estimate), with an additional step to debias the power spectrum. This removes most of the contamination present, but can be simply adapted to, and significantly improved by, treating the multiplicative component of contamination instead of just the additive term. We demonstrate how to do this in Sec.~\ref{sec:addmultmethods}. In all cases we studied, the error from not correcting the multiplicative term dominated over error induced from overfitting --- as Fig.~\ref{fig:nclean} illustrates, in the \textit{ideal} case where our templates exactly matched the systematics, not treating the multiplicative term introduced as much error as using $\sim30\times$ more templates than systematics in the cleaning procedure. 
    \item \textbf{Template Subtraction:} Equivalent to performing an individual OLS regression at each multipole, resulting in large variance and significant loss of signal from overfitting. As a result, it does not reconstruct maps well and generally performs most poorly in all of our tests. However, our implementation is a limiting case, where each harmonic from each template is allowed to contaminate independently, in contrast to Mode Projection where all modes contaminate identically. The work here should make it straightforward to construct a hybrid method where all modes contribute identically like in Mode Projection (as is physically motivated) and hence have small variance, but where certain modes are prioritized for cleaning, based on the analysis case. 
    \item \textbf{Iterative Forward Selection:} This is a method we propose, which is a much simpler version of the DES-Y1 method that requires only a single tunable parameter (a significance threshold) and no mocks. We found that it produces excellent results and is robust to correlation between templates, but is not as robust to overfitting, displaying the same dependence of roughly $\dchisqcl\propto\Ntpl$ as the Mode Projection and Template Subtraction methods. 
    \item \textbf{Cross-Validated Elastic Net}: A cleaning method we introduce, which we find has the best overall performance, being consistently low error and robust to overfitting. It is equivalent to Mode Projection, but with the amplitude of contamination for each template having a mixed Gaussian/Laplace prior applied to encourage sparsity and thus automatically select the important templates. The `priors' are not strictly such in a Bayesian sense, as their strengths are determined by the data through cross-validation. It is easy to implement using out-of-the-box software and doesn't require a user-defined prior for the power spectrum or debiasing step, providing the best balance of performance, ease of implementation, interpretability and robustness.
\end{itemize}

\section{Conclusions}\label{sec:conclusion}
In this paper, we carried out a broad comparison of methods used to remove astrophysical, atmospheric, and instrumental systematic errors that affect galaxy-clustering measurements.
We have generalized previous work by 1) showing how different methods can be interpreted under a common regression framework, 2) jointly assessing the robustness of methods on simulated data, 3) investigating the reconstruction fidelity of LSS \textit{map(s)}, rather than just their clustering statistics, as the maps are useful points of departure for numerous other analyses (e.g.\ summary statistics beyond the power spectrum, cross-correlations, searches for signatures of dark matter or exotic new physics); and 4) proposing improvements to current methods, as well as new, hybrid and efficient methods for the systematics cleaning.

We employed a simple and general model for systematics, given in Eq.~(\ref{eq:model}), which allows for spatially varying multiplicative and additive systematic errors with a range of clustering properties to any generic cosmological field. Equipped with that model, we defined a testing procedure that attempts to mimic real-world conditions for LSS  surveys, where the true galaxy map is contaminated with an unknown set of systematics and a set of known templates is used to model and correct for the contamination. Given our methodology (pictorially described in Fig.~\ref{fig:flowchart}) and a set of assumptions about the fiducial DES Y5-like survey used to generate the maps, we studied the performance of the  systematics-cleaning methods under different conditions.

We showed that both Template Subtraction and Mode Projection, while developed independently, can be interpreted through a regression framework where the signal of interest corresponds to the noise term of a regression model. This allowed us to straightforwardly apply known statistical results and techniques to these methods. We used this to adapt additive methods to account for multiplicative errors (Fig.~\ref{fig:pixrmsedist}), and identify potentially highly contaminated map pixels as a function of their ``leverage" (Fig.~\ref{fig:pixerrhii}), while opening up avenues for further improvement. One such avenue we touched on was to optimize Mode Projection (or other regression methods) by prewhitening the maps in harmonic space. Recognizing that the noise of the regression is the clustering signal itself, we proposed that the maps could be efficiently and optimally inverse-variance weighted in harmonic space, where the clustering signal is diagonal. This is equivalent to accounting for the off-diagonal pixel covariance in the pixel-based regression methods, which is rarely done for tractability reasons (but see \citet{wagoner_des_sys} for one approach). We found this to improve results (Fig.~\ref{fig:prewhiten}), but be subdominant to the multiplicative correction and differences between the cleaning methods. 

We introduced two new methods for cleaning: (1) the `Forward Selection' method, which is a greatly simplified version of the DES-Y1 method that achieves similar performance albeit being less robust to a large number of templates; and (2) the `Elastic Net' method, a simple out-of-the-box method that implements Mode Projection, but which automatically selects important templates. We found that the Elastic Net method is very robust, with strong performance even when there is a large number of templates (Fig.~\ref{fig:nclean}) or templates  are highly correlated (Fig.~\ref{fig:ncleanxcorr}); both are cases where other methods display weaknesses. This method is very easy to implement, and we recommend it for future surveys.

On the whole, we found that all of the methods perform quite well, dramatically improving the chi-squared difference between the cleaned and true (uncontaminated) angular power spectrum. At the map level, Template Subtraction was the only method that did not significantly reduce the RMS overdensity error across pixels (Figs.~\ref{fig:pixerrhii} and \ref{fig:nclean_rmsmap}), and so we do not recommend the version implemented here for map reconstruction. Once we adopted only the algorithms that take into account both additive and multiplicative errors, all of the methods improved $\Delta\chi^2_{C_\ell}$ by three orders of magnitude relative to the uncleaned case. Moreover, overfitting did not lead to large degradation in the reconstructed power spectra (see Fig.~\ref{fig:nclean}), which is encouraging. Finally we found that the performance of the various systematics-cleaning methods is very weakly dependent on the level of cross-correlation between the template maps used for the cleaning, with the DES-Y1 method being mildly more susceptible.

We end with several recommendations based on this work:

\begin{enumerate}[I.]
\item Current and future cleaning methods should account for multiplicative contamination. `Weights' methods like the DES-Y1 method already do this and other methods like (Pseudo-$\Cl$) Mode Projection can easily do so via Eqs.~(\ref{eq:debiasedols2})-(\ref{eq:debiasedols22}).
\item Cleaning methods based on a single Ordinary Least Squares regression are equivalent to (Pseudo-$\Cl$) Mode Projection and so should debias inferred two-point functions accordingly. 
\item Analyses should err on the side of overfitting rather than underfitting for templates, as the error from the former tends to be small. Researchers should avoid arbitrarily removing templates from the library prior to cleaning based solely on their similarity to other templates. Larger template libraries result in increased variance of the map and power spectrum estimators, especially with the very large number of templates that will be available to future surveys. Therefore:
\item In scenarios where a very large template library is available, the data itself should be used to select a subset for cleaning. Among the methods that we studied, this is accomplished by either a DES-Y1 type method or the Elastic Net with cross validation. Both show good robustness, and the latter is simple to implement with common software. The theoretical connections we have made between methods should make alternative template selection routines such as those in Refs.~\citep{Leistedt:2015kka, Rezaie:2019vlz} simple to adapt and implement.
\item The cleaning methods used thus far can --- and should --- be viewed in the context of regression, with the estimated overdensity field corresponding to the regression residuals. Researchers should make use of the powerful suite of existing tools and diagnostic measures to assess the validity of regression models when cleaning LSS data (e.g. leverage for outlier detection, Q-Q plots, partial regression/residual plots) and to aid mask creation. This is applicable to all methods studied in this paper.
\end{enumerate}

\acknowledgments We thank Boris Leistedt and Jessie Muir for helpful comments and encouragement. The authors have been supported by DOE under Contract No. DE-FG02-95ER40899. D. H. has also been supported by NASA under Contract No. 14-ATP14-0005, and N. W. has been supported through a Leinweber Graduate Fellowship.


\appendix

\section{Lognormal vs. Gaussian Signal Maps}\label{app:lognorm}
While the methods presented here are quite general for any case where systematic contamination can be traced using a template, we have specifically worked in the context of galaxy clustering. In this case, the signal map $s$ that we are attempting to model is the galaxy overdensity $\delta$, which is subject to the constraint $\delta > -1$ (as is the case for any overdensity statistic). Thus our assumption that $s$ is Gaussian breaks down at low redshift and at small scales, when $|\delta|$ can be large.

\begin{figure}[htbp]
\includegraphics[width=\linewidth]{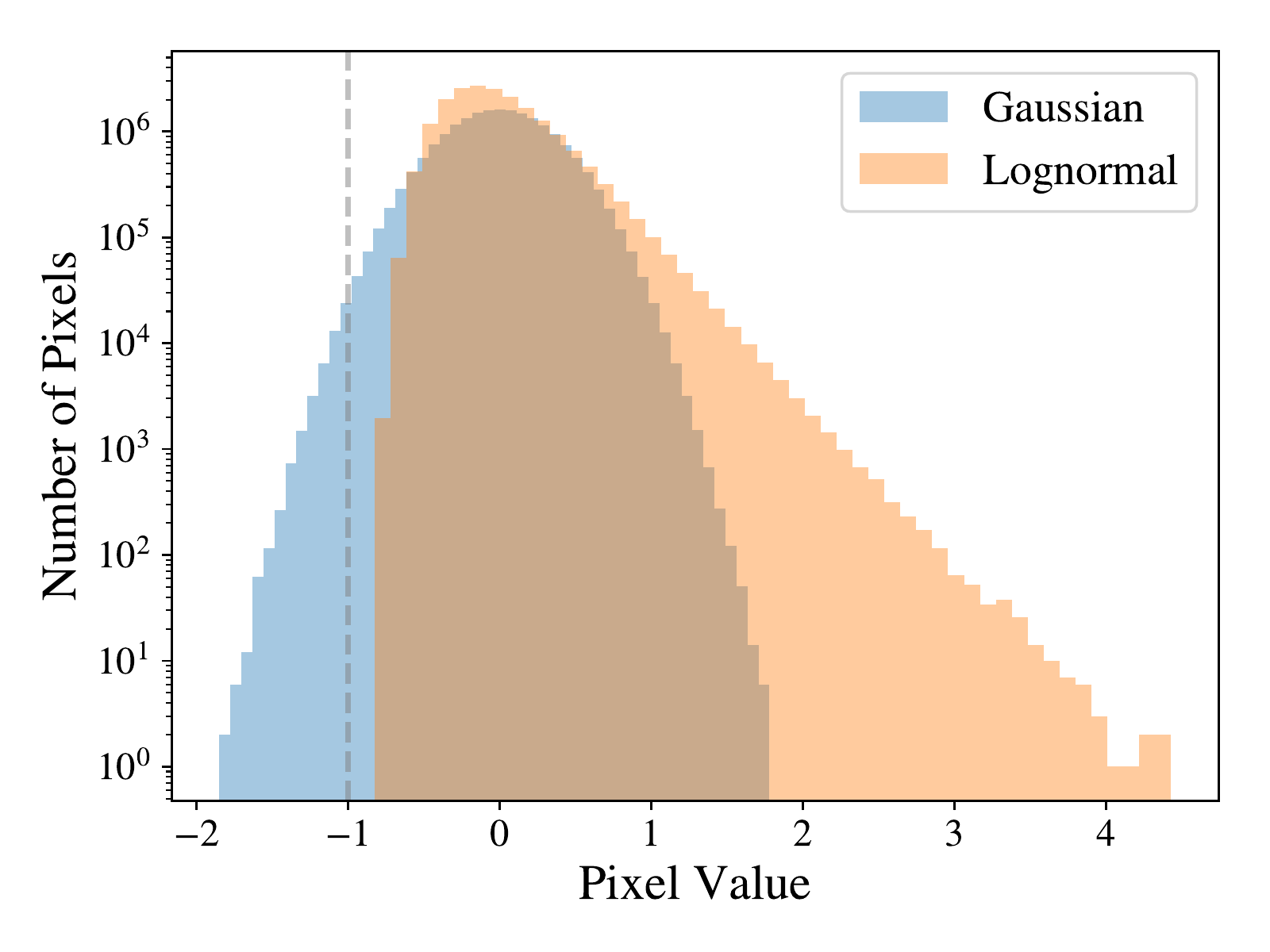}
\caption{Distribution of pixel overdensities across all 100 realizations of the lognormal (orange) and Gaussian (blue) maps of the galaxy overdensity in the lowest redshift bin of our fiducial DES-like survey. The Gaussian maps contain pixels with $s<-1$, which is nonphysical in cases like this where $s$ corresponds to an overdensity.}
\label{fig:lognormpixdist}
\end{figure}
It is well known that galaxy and shear overdensities are better approximated by a lognormal distribution (see e.g. Refs.~\cite{coles_lognormal_1991, Taruya_2002, Hilbert_2011, Xavier_2016}), so we run the methods on a series of lognormal maps to see if the relative performance of the methods changes.

We generate 100 Gaussian signal realizations $s_G(\nhat)$ of the lowest redshift bin of our fiducial DES survey, for which the cosmological signal will be most non-Gaussian. We generate lognormal versions of these maps by first computing the transformation that achieves zero-mean lognormal overdensity field \textit{in the ensemble} \cite{Hilbert_2011}, then centering and scaling so that each realization of the lognormal field has the same mean and variance as its Gaussian counterpart. The two steps correspond to the mathematical operations:
\begin{enumerate}
    \item $s'_{LN}(\nhat) = e^{s_G(\nhat)} - e^{{\rm Var}[s_G(\nhat)]/2}$ \\
    \item $s_{LN}(\nhat) = \sqrt{\frac{{\rm Var}[s_G(\nhat)]}{{\rm Var}[s'_{LN}(\nhat)]}} \left(s'_{LN}(\nhat) -  \bar{s'}_{LN}(\nhat)\right)$.
\end{enumerate}

The resulting lognormal realizations are then of the form
\be
s_{LN}(\nhat) = \lambda_1 e^{s_G}(\nhat) - \lambda_0,
\ee
with scale and shift parameters of $\lambda_1=0.9123\pm0.0017$ and $\lambda_0 = 0.9697\pm0.0017$, respectively for our lowest redshift bin, which is the most non-Gaussian. 

Fig.~\ref{fig:lognormpixdist} shows the distribution of pixel overdensities across all realizations of the lognormal and Gaussian signal maps. It is clear that the Gaussian maps contain many pixels with $s<-1$, which is nonphysical for our case, where $s$ corresponds to an overdensity. The lognormal maps avoid this problem and are a better approximation of the true overdensity distribution. 
As we have shown, most of the cleaning methods can be viewed under a regression framework wherein the signal distribution is assumed to be Gaussian, so we investigate whether our comparison of methods changes when using a more realistic lognormal distribution.

\begin{figure}[htbp]
\includegraphics[width=\linewidth]{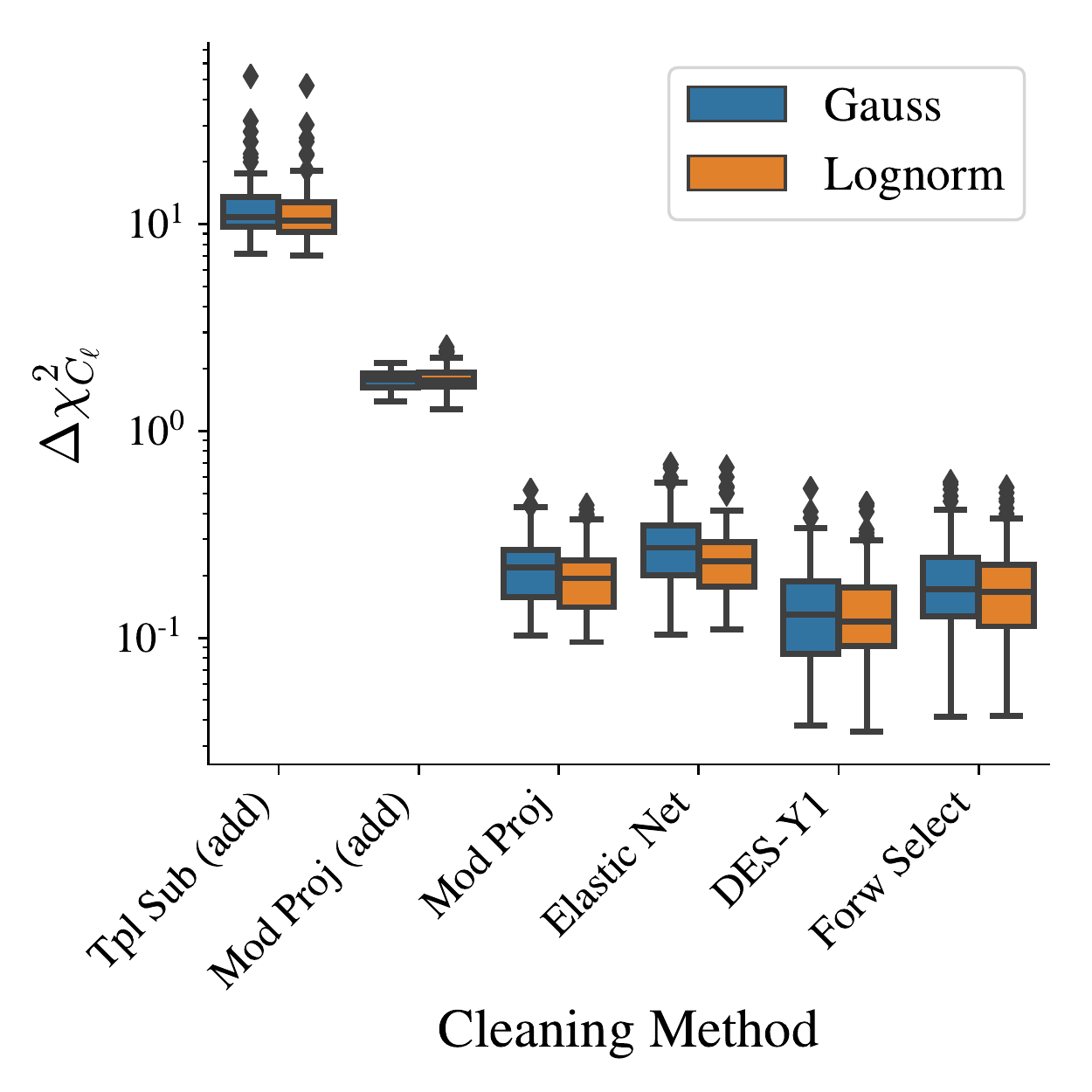}
\caption{Box plot showing the performance of each cleaning method when using Gaussian (blue, left) versus lognormal (orange, right) signal maps, as measured by $\dchisqcl$ of the power spectrum. Filled boxes show the 25-50-75\% quartiles, with whiskers encompassing the rest of the distribution out to $1.5\times$ the inter-quartile range. Points beyond this range are indicated by diamonds. Regardless of whether lognormal or Gaussian maps are used, the relative performance of the methods to one another is largely unchanged, and the Gaussian approximation is negligible compared to neglecting the multiplicative correction of Sec.~\ref{sec:addmultmethods}.}
\label{fig:lognorm-boxplot}
\end{figure}

Fig.~\ref{fig:lognorm-boxplot} shows the error in the power spectrum reconstruction, given by the $\Delta\chi^2_{C_\ell}$ statistic, for  the different methods. We find that while there is some overall shift, using the lognormal signal maps does not change the relative behavior of the methods; none of them display a unique susceptibility to the assumption of Gaussianity in the signal maps.

\section{Effect of Prewhitening}\label{app:prewhitening}

\begin{figure}[htbp]
\includegraphics[width=\linewidth]{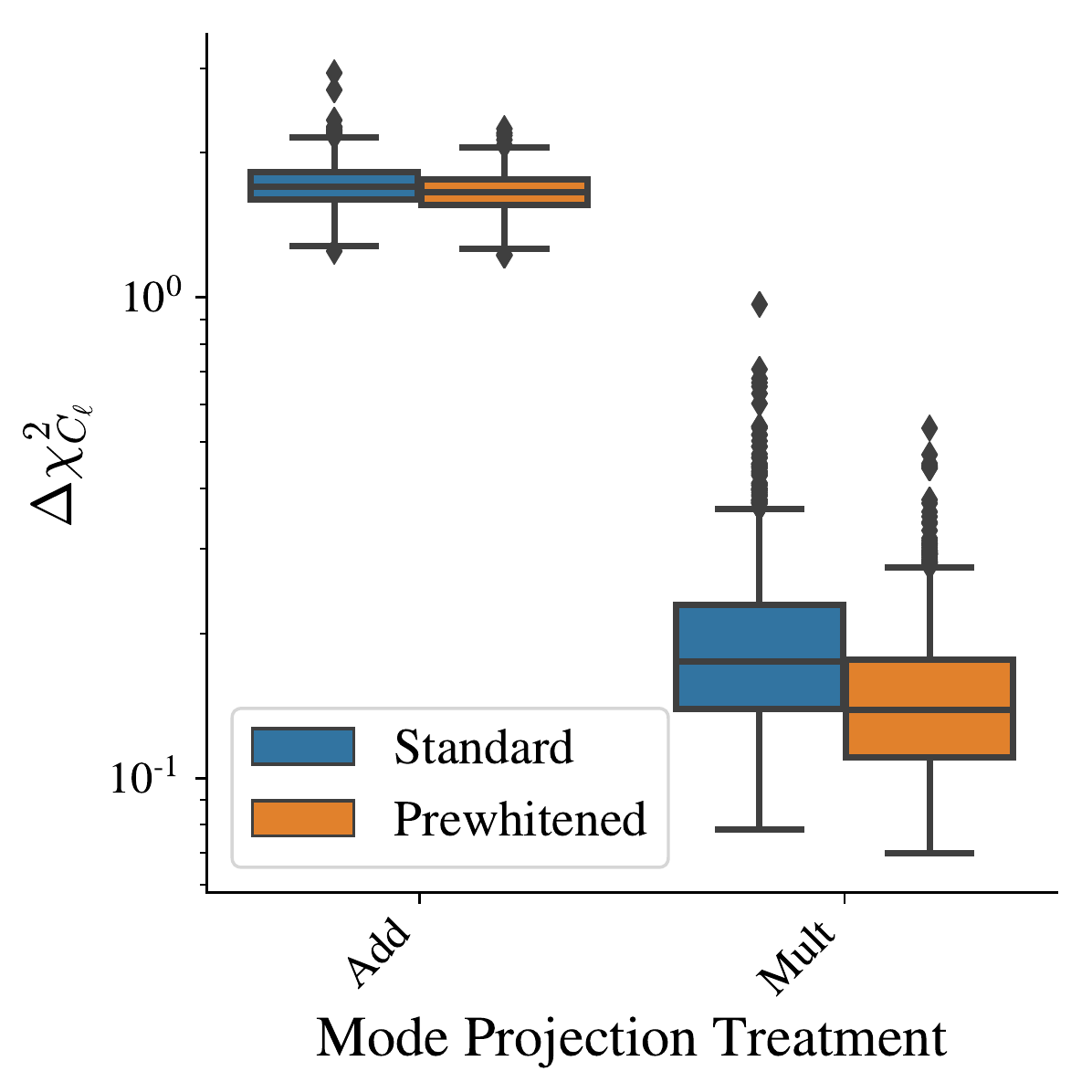}
\caption{Impact of prewhitening before cleaning with the multiplicative and additive versions of the Mode Projection method on 1000 realizations for our fiducial contamination model. The standard Mode Projection method assumes a flat power spectrum for the target signal, resulting in a suboptimal estimate of contamination. This can be improved through `prewhitening' the data vector and templates using a prior power spectrum, which can be shown to be equivalent to a standard weighted regression procedure in harmonic space. There is clear but modest improvement from the standard case (blue) to the nearly-optimal, prewhitened case (orange), with the most improvement seen for realizations that have large error. This can be seen by the preferential reduction of extreme points at the high end of the box plots in the prewhitened case (note the log scale).}
\label{fig:prewhiten}
\end{figure}

In their derivation of the bias on the estimated power spectrum after (pseudo-$\Cl$) Mode Projection, \citet{Elsner:2016bvs} assume that the map $d$ has been decorrelated (``prewhitened") before projecting out the templates. This is quite difficult to do in practice, as it requires the inversion of an $\Npix \times \Npix$ matrix, the same problem with QML estimators for the power spectrum. Indeed, one of the assumptions of pseudo-$\Cl$ estimation is that pixels are uncorrelated (though individual pixels are weighted by an estimate of their inverse noise variance and by the mask, see e.g. Ref.~\cite{Alonso:2018jzx}.) 

As shown in Sec.~\ref{sec:regression}, however, the dominant `noise' in our observations is actually our true clustering signal, so a true `prewhitening' step should more appropriately inverse weight the data by the expected clustering variance. This can be done efficiently in harmonic space when there is no mask, as the clustering signal is diagonal, circumventing the need to invert a large covariance matrix. 

We can define prewhitened data vectors for our observed overdensity field and templates as
\be
(\dobs)_{\ell m}' = (\dobs)_{\ell m}/\sqrt{\Clss}.
\ee
\be
(t_i)_{\ell m}' = (t_i)_{\ell m}/\sqrt{\Clss},\label{eq:tplprewhiten}
\ee
which results in coefficient estimates of 
\begin{align}
\mathbf{\hat\alpha} &= (\mathbf{T'}^{\,\dag} \mathbf{T'})^{-1} \mathbf{T'}^{\,\dag} \mathbf{\dobs'}, 
\end{align}
where $T'$ is a $N_{\ell m} \times \Ntpl$ matrix with complex entries defined in Eq.~(\ref{eq:tplprewhiten}).
We can compute the amplitudes directly with
\be 
\hat{\alpha} = \frac{\sum_{\ell=0}^{\ellmax} (2\ell + 1)\tilde{\Cl}^{td}/\Clss}{\sum_{\ell=0}^{\ellmax} (2\ell + 1)\tilde{\Cl}^{tt}/\Clss}.
\ee

We found that prewhitening improved  $\dchisqcl$ by a mean of $\sim0.05$ with dispersion $0.08$ across the mocks, with similar shifts regardless of whether the multiplicative correction was applied or not. Fig.~\ref{fig:prewhiten} shows the improvement from the standard case (blue) to the prewhitened case (orange) for both additive and multiplicative mode projection. While we do not show it, we found that the benefit of prewhitening increased for realizations that had worse power spectrum estimates (higher $\dchisqcl$), in effect catching and mitigating particularly bad realizations. 
 
In practice, one would either assume a prior power spectrum for prewhitening or compute it iteratively, just as one does for the Mode Projection debiasing step, so this could easily be incorporated into existing Mode Projection routines such as \texttt{NaMaster}. As noted in Sec.~\ref{sec:regression}, since Mode Projection is equivalent to regression, this improvement also quantifies the expected level of improvement that would come from accounting for the covariance between pixels in pixel-based regression methods.

\begin{figure*}
    \includegraphics[width=0.48\textwidth]{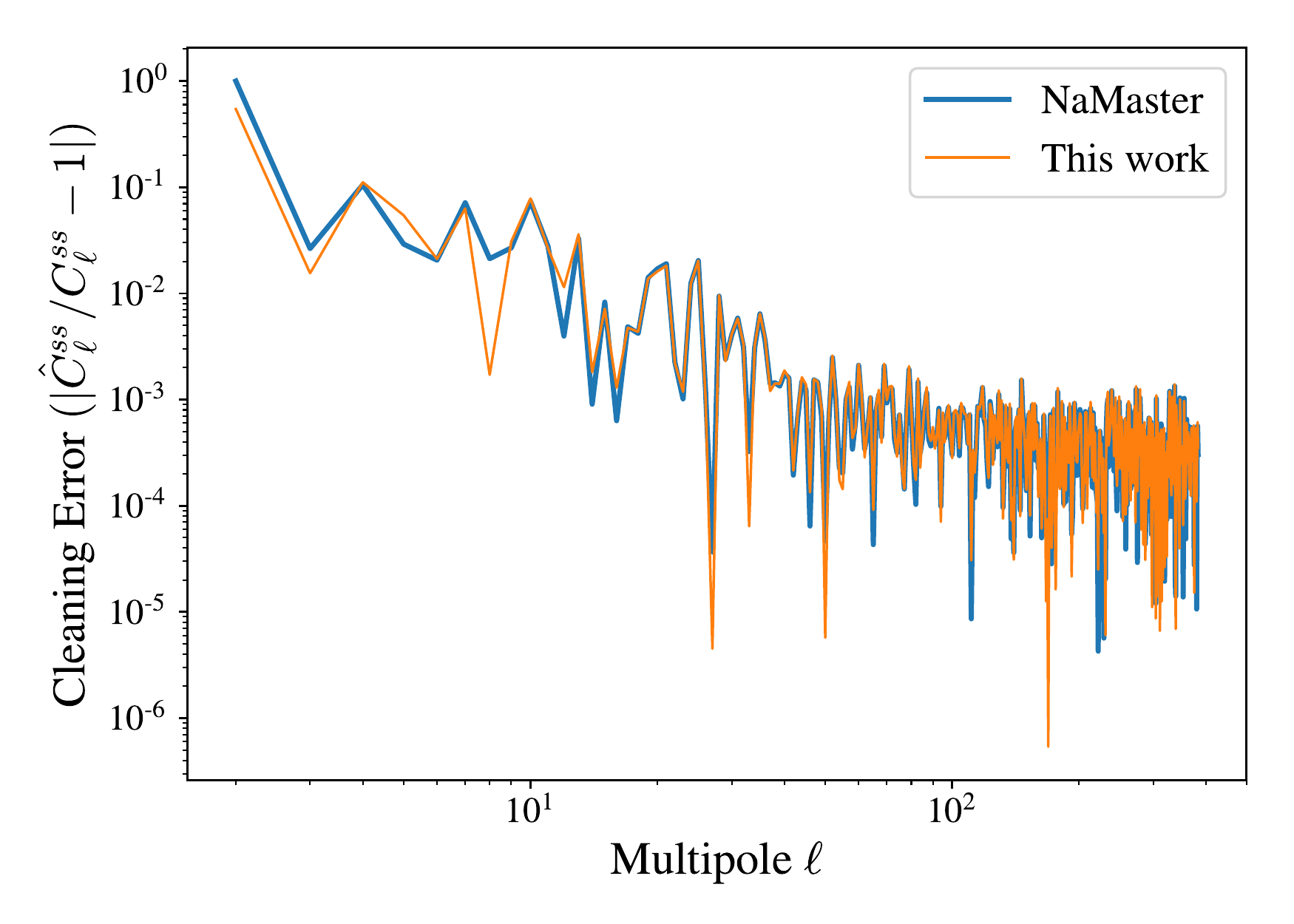} 
    \includegraphics[width=0.48\textwidth]{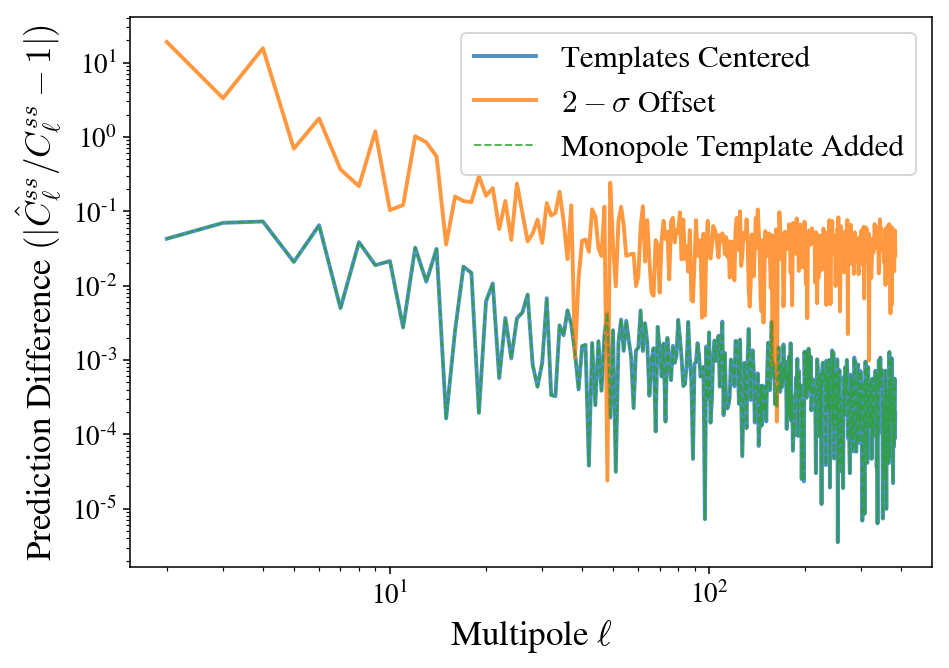} 
        \caption{Validation tests of the Mode Projection map-cleaning procedure. \textit{Left panel:} Comparison of Mode Projection performance on an additive-only contaminated map (as assumed by the Mode Projection method), using \texttt{NaMaster} (blue) and our own implementation (orange). The agreement between the two is very good. \textit{Right panel:} Impact of not pre-centering cleaning templates in \texttt{NaMaster}. The blue curve indicates the standard use case, where contamination is additive and completely described by the templates, which have been individually centered at zero. If templates are instead centered at another value (here we add a constant $2\sigma_{{\rm tpl}, i}$ offset to each template, where $\sigma_{{\rm tpl}, i}$ is the standard deviation of values in template map $i$. Adding a monopole template completely mitigates the bias from non-centered templates.}\label{fig:namaster}
\end{figure*}

Analyses on real data will of course be complicated by the mask, which correlates different multipoles, but this can be addressed by suitable binning of the multipoles. Indeed, the standard pseudo-$C_\ell$ Mode Projection assumes a flat power spectrum and so can be thought of as the limiting case of using only a single bin across multipoles with equal weighting, such that even a rough estimate of the signal power spectrum should offer improvement.

The other methods tested here should benefit similarly from prewhitening, with the possible exception of the DES-Y1 method, which already incorporates an estimate of the covariance of $s$ (which accounts for much of the methods' complexity). The Forward Selection method we presented may be particularly impacted, since the estimated covariance of the fit parameters is underestimated when the pixel covariance is neglected, and this is used for the significance criterion for selecting a template. This could be one reason why the Forward Selection method sometimes failed to reduce all templates to below a significance of $\dchisqthresh=2$ --- such a threshold was artificially low compared to what would be expected from random variation.  

As noted in Eq.~\ref{eq:syscov}, the prewhitening step in Eq.~(\ref{eq:tplprewhiten}) should optimally include contributions from the systematics as well. However as this represents minor perturbations to the major prewhitening correction above and is hence a small `error on the error', the effects should be small. This is consistent with \citet{Elvin-Poole:2017xsf}, who found negligible impact on their method from neglecting the additional systematics contribution to their estimated covariance matrices.

\section{Comparison with \texttt{NaMaster}}\label{app:namaster}

We have used our own implementation of the Mode Projection method and have tested it against that of \texttt{NaMaster} \cite{Alonso:2018jzx}, finding good agreement. \texttt{NaMaster} computes the power spectra given a set of templates and observations, but does not produce map estimates, so we compare the two implementations using the cleaned power spectrum only. The left panel of Fig.~\ref{fig:namaster} shows the relative error of the estimated power spectrum when cleaned using NaMaster vs. our own implementation, using the exact same contaminated map and templates and we find good agreement (this held true for all realizations tested). There is very slight disagreement at larger scales (low $\ell$), which may be numerical artifacts from the \texttt{Master} \cite{Hivon:2001jp} algorithm implemented to account for mode coupling on a cut sky being applied to full-sky input maps. Regardless, the deviations between the two are small for $\ell>2$. 

\section{Accounting for the Monopole}\label{app:monopole}

It is worth saying a few words about the monopole term, both as in terms of prediction and as it relates to regression.

Firstly, the \textit{overdensity} residuals do not correspond to the \textit{number} density residuals. Even with a perfect reconstruction $\shat = s$, the true number density will be unknown up to a factor of $\gamma$,
\be
\Ntrue = \gamma\pixavg{\Nobs}(s+1),
\ee
and as such the estimated number density could be quite different from the truth. Fig.~\ref{fig:add_mult_resids} shows a somewhat unintuitive consequence of this. The single sytematic that contaminates the field has the form $\fsys \propto -t$, so that it only obscures galaxies from view ($\fsys \le 0$). 
At $t=0$, there is no contamination and so $\Nobs=\Ntrue$, however as the figure shows the \textit{over}-density residuals are quite large.
This is because the \textit{mean} number density is significantly underestimated, so pixels with no obscuration are preferentially (and wrongly) estimated to reside in overdense regions.\footnote{In other words, \ $\gamma > 1$, so from Eq.~(\ref{eq:dobsgamma}), $\pixavg{\dobs\rvert_{\fsys=0}} > 0$).} 

Secondly, a net monopole in $\fsys$ corresponds to the intercept in the regression methods (a column of ones in $T$). 

In OLS regression (or pseudo-$C_\ell$ Mode Projection), the fit is guaranteed to go through the center of mass of the points, $(\bar{t}, \bar{d}_{\rm obs})$, such that including a monopole is unnecessary with such methods if working with overdensities and zero-centered templates. In such cases, the `projection' of the monopole has already been done by subtracting the mean from the density and template maps (consider Eq.~(\ref{eq:sestmp}) with a template of all 1s). We showed in Eqs.~(\ref{eq:dobsgamma})-(\ref{eq:hatfsysadd}) how how this also holds in the multiplicative case.

In realistic situations, there is high susceptibility to human error if a monopole term is not included --- previously zero-centered maps can easily shift through template transformations, mask adjustments, and the application of a mask to mocks, resulting in wildly biased contamination estimates that may be difficult to detect. For example, it is easy to pass templates that are not zero-centered to current pseudo-$\Cl$ Mode Projection methods such as implemented in \texttt{NaMaster} and receive highly biased spectra without warning (see right panel of Fig.~\ref{fig:namaster}).

The DES-Y1 and Forward Selection methods both already include an intercept term, in keeping with the original formulation of the DES-Y1 method, though in practice it should be very close to zero.

We therefore opt to include a monopole term in our Elastic Net method, as this ensures the method is robust and generalizes the process beyond overdensities to non-zero mean fields, and it will naturally be ignored as a template if it does not contribute information.

\begin{figure*}[htbp]
\includegraphics[width=\linewidth]{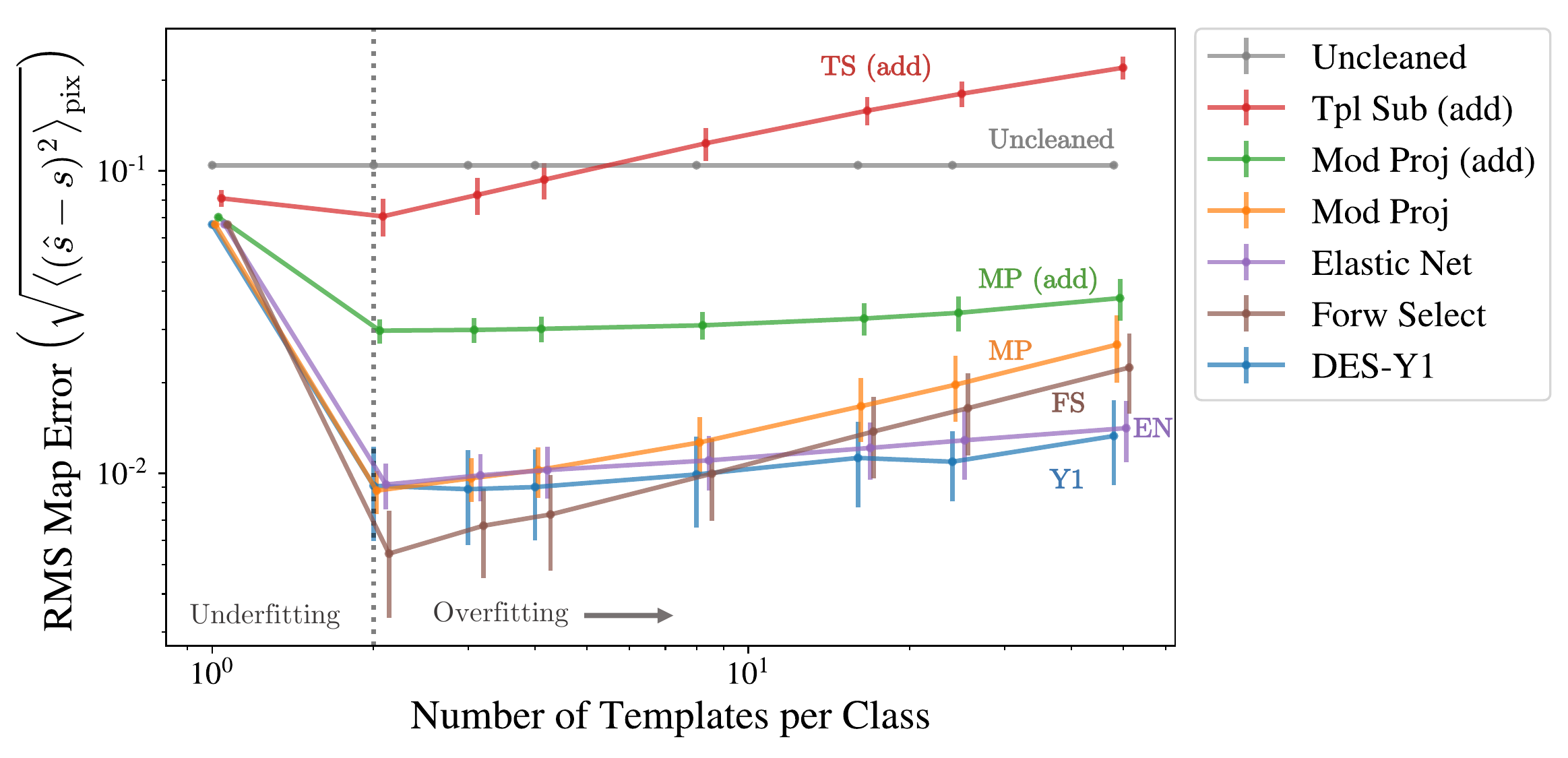}
\caption{Same as Fig.~\ref{fig:nclean} but showing the RMSE in the estimated overdensity map for each method, rather than error in the power spectrum. Trends are very similar. See Sec.~\ref{sec:overfit} for details.}
\label{fig:nclean_rmsmap}
\end{figure*}

\section{Impact of $\dchisqthresh$ on DES-Y1 Analysis}\label{app:Y1}

Here we investigate the effect of $\dchisqthresh$ and $\varsys$ on the efficacy of the DES-Y1 method, as described in Sec.~\ref{sec:Y1}. We describe the reconstruction quality with the residual chi squared between the cleaned and true model, $\dchisqcl$.

\begin{figure}[htbp]
\includegraphics[width=\linewidth]{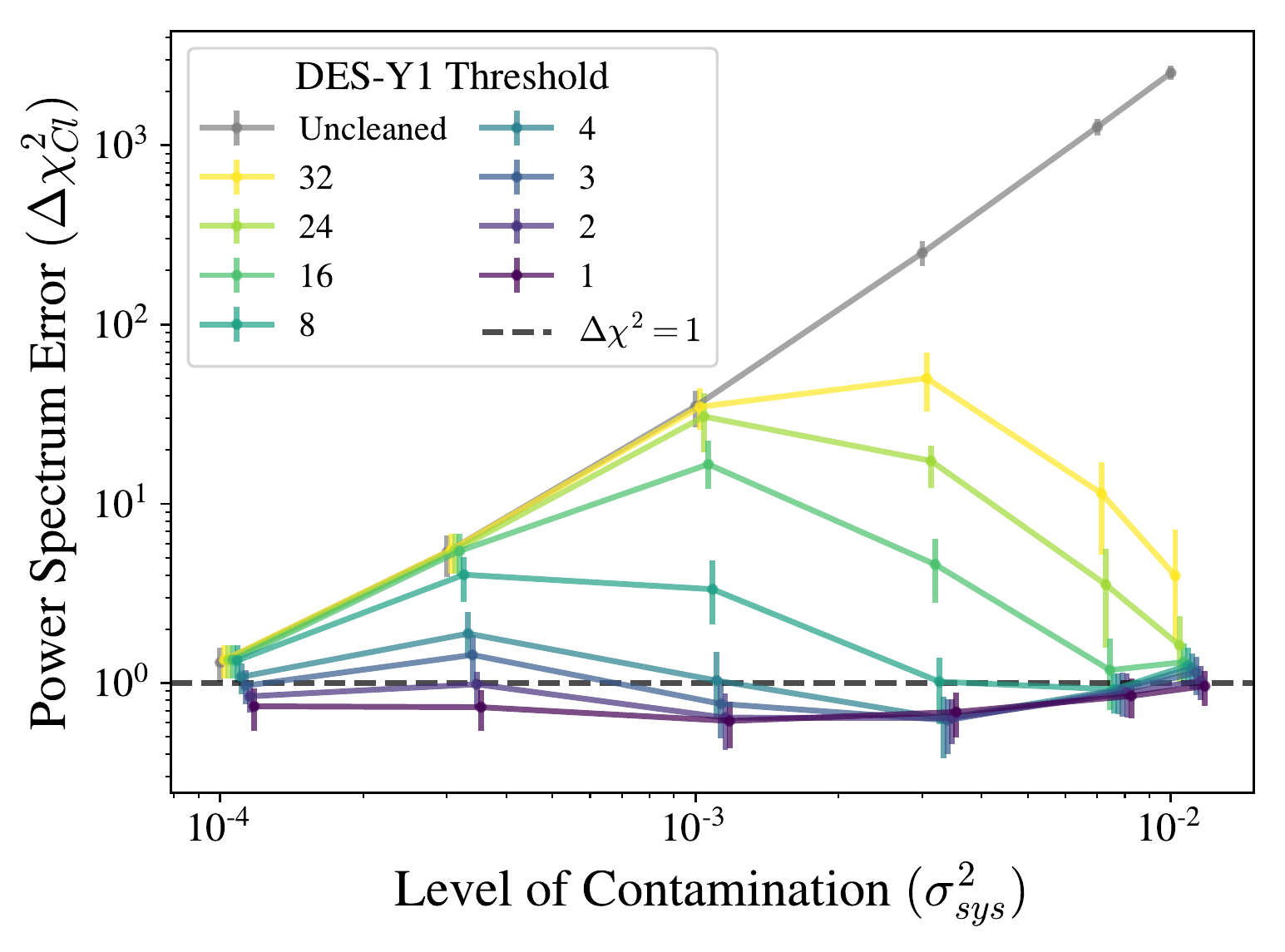}
\caption{Dependence of the power spectrum error ($\dchisqcl$) on the level of contamination $\varsys$ ($x$-axis), and on the stopping criterion 
$\dchisqthresh$ used for the DES-Y1 method (colors). Points are offset for clarity. For comparison, the variance across pixels from the true overdensity in each bin ranges from $\sigvar \in [0.075, 0.122]$ for the 5 redshift bins of our fiducial survey, corresponding to factors of 7.5 --- 1220$\times$ larger than $\varsys$ for the points shown.}
\label{fig:epthreshsys}
\end{figure}

Fig.~\ref{fig:epthreshsys} shows how $\dchisqthresh$ affects the reconstruction quality for the DES-Y1 method, as a function of the level of contamination parameterized by the systematic-error variance $\varsys$. We find little reduction in error by lowering the significance threshold below $\Delta\chi^2_{\rm threshold} = 4$. 

At our fiducial level of contamination ($\varsys = 10^{-2}$), almost all contaminating templates exceed the highest threshold displayed of $\dchisqthresh = 32$ and so are corrected for. The larger the contamination, the more precisely its form can be determined, so as the level of contamination decreases, some contaminated templates are left uncorrected for. This results in the somewhat counter-intuitive turnover in the error for a given threshold level. We found that the lowest threshold of $\dchisqthresh=1$ consistently outperformed higher thresholds, despite the risk of overfitting, in agreement with our results in Sec.~\ref{sec:overfit}, which showed that the extra power from residual contamination is likely more pernicious than the excess removal of power due to overfitting.

-----------------------
\bibliography{paper_ggsys}{}
\end{document}